\documentclass{article}

\usepackage{arxiv}
\usepackage{lipsum}   
\usepackage{marvosym}
\usepackage{adjustbox}
\usepackage[inline]{enumitem}
\usepackage{graphicx} 
\usepackage{tikz}
\usepackage{tikz-qtree}
\usetikzlibrary{trees} 
\usepackage{amsmath} 
\usepackage{amssymb}  
\usepackage{color}
\usepackage[ruled,vlined,linesnumbered]{algorithm2e}
\usepackage{float}
\usepackage{caption}
\usepackage{subcaption}
\usepackage{hvfloat}
\usepackage{mathtools}
\usepackage{multirow}
\usepackage[title]{appendix}
\usepackage[normalem]{ulem}
\usepackage{xcolor}
\usepackage{colortbl}
\usepackage{epigraph}
\usepackage{bbm}
\usepackage{comment}
\usepackage{authblk}
\usepackage{etoolbox}
\usepackage{rotating}
\usepackage{tabularx}
\usepackage{empheq}
\usepackage{pifont}
\usepackage{makecell}
\usepackage{setspace}

\newtheorem{definition}{Definition}[section]

\DeclareMathOperator*{\argmin}{arg\,min} 
\usepackage[utf8]{inputenc} 
\usepackage[T1]{fontenc}    
\usepackage{hyperref}       
\usepackage{url}            
\usepackage{booktabs}       
\usepackage{amsfonts}       
\usepackage{nicefrac}       
\usepackage{microtype}      
\usepackage{lipsum}
\usepackage{graphicx}
\graphicspath{ {./images/} }

\title{Empirical Evaluation of Structured Synthetic Data Privacy Metrics: Novel experimental framework}

\author[1,2,*]{Milton Nicol\'as Plasencia Palacios}
\author[1]{Alexander Boudewijn}
\author[1]{Sebastiano Saccani}
\author[3]{Andrea Filippo Ferraris}
\author[1]{Diana Sofronieva}
\author[4]{Giuseppe D'Acquisto}
\author[4]{Filiberto Brozzetti}
\author[1]{Daniele Panfilo}
\author[2]{Luca Bortolussi}

\affil[1]{Aindo SpA, Trieste, Italy}
\affil[2]{University of Trieste (UniTs)}
\affil[3]{LAST-JD, Alma AI, Alma Mater Studiorum, University of Bologna \& DIKE research group, PREC department, Vrije Universiteit Brussel}
\affil[4]{Luiss University, Rome, Italy}
\affil[*]{Corresponding author, plasencia.milton@gmail.com}

\begin{document}
\maketitle
\begin{abstract}
Synthetic data generation is gaining traction as a privacy enhancing technology (PET). When properly generated, synthetic data preserve the analytic utility of real data while avoiding the retention of information that would allow the identification of specific individuals. However, the concept of data privacy remains elusive, making it challenging for practitioners to evaluate and benchmark the degree of privacy protection offered by synthetic data. In this paper, we propose a framework to empirically assess the efficacy of tabular synthetic data privacy quantification methods through controlled, deliberate risk insertion. To demonstrate this framework, we survey existing approaches to synthetic data privacy quantification and the related legal theory. We then apply the framework to the main privacy quantification methods with no-box threat models on publicly available datasets. \let\thefootnote\relax\footnote{Funded by the European Union though grant agreement 101218531 — SydAi}
\addtocounter{footnote}{-1}\let\thefootnote\svthefootnote
\end{abstract}

\keywords{synthetic data \and privacy \and privacy quantification\and attacks on privacy \and privacy measurements \and machine learning and privacy}

\section{Introduction}

Synthetic data generation is increasingly recognized as a powerful privacy-enhancing technology (PET). By preserving the properties of real-world data, synthetic data allow organizations to mitigate privacy risks in research, innovation, and the development of artificial intelligence (AI) systems. This has led to a growing interest in the PET in high-impact sectors such as healthcare~\cite{Giu2023heatlh, Goncalves} and finance~\cite{potluru2024finance}. 

However, there is no universal standard for quantifying privacy risks, leading to fragmented approaches in practice. Privacy is an interdisciplinary concept that spans legal, technical, and ethical dimensions. Practitioners often struggle to justify their choice of privacy metrics or methods, given the lack of clear guidance and the various tools available. Moreover, the practical challenges of meeting regulatory compliance further complicate this task and can limit innovation in data-driven fields \cite{GalLynskey2024}. This highlights the urgent need for comprehensive analyses and standardized frameworks for privacy assessment, as identified in prior works~\cite{reiter2023synthetic, Exeter, Bellovin18, raab2024privacymetrics, Boudewijn}.\\
\indent To address these challenges, this paper introduces a benchmarking framework for privacy quantification methods in the context of synthetic data generation. This framework is centered around deliberate and controlled insertion of risk to synthetic datasets in both an idealized manner (directly leaking proportions of real records, based on~\cite{Anonymeter}), and in manners modeling how privacy risks occur in practice (generator overfitting and privacy budgets). To demonstrate the framework, we first survey existing methods for quantifying synthetic data privacy. We then describe our benchmarking framework and apply it in an experimental analysis of the surveyed privacy quantification approaches. Throughout, we relate our framework, as well as privacy metrics, to relevant legal standards and theory.\\
\indent Our findings indicate that most metrics consistently reflect controlled risks in both theoretical and practical scenarios. Measured risk is typically proportional to inserted risk in idealized settings, though data characteristics, such as a high proportion of categorical attributes, can amplify privacy risks. By combining legal insights, a comprehensive classification of privacy quantification methods under a no-box threat model, and a robust benchmarking framework, our work contributes to the standardization of privacy assessment for synthetic data. This supports researchers and practitioners in advancing synthetic data as a viable PET in real-world applications.

\section{Background}
\subsection{Synthetic data}



Contemporary usage of synthetic data has become increasingly significant in healthcare \cite{Gonzales2023, Giuffre2023}, finance, and research, owing to its notable privacy-enhancing advantages and capacity to foster data-driven innovation. Synthetic data is often hailed as a promising privacy-enhancing technology \cite{DBLP, Ebert, avatar, RAJOTTE2022105331}. However, apprehensions about regulatory sanctions frequently deter market actors from fully adopting synthetic data as its status within the regulatory framework remains ambiguous \cite{GalLynskey2024, Boudewijn}, where standardized assessment methodologies and clear legal criteria are lacking \cite{Boudewijn2024}.

Following Jordon et al.~\cite{jordon2022synthetic}, we formalize the concept of synthetic data through Definition~\ref{def:SD}.

\begin{definition}\label{def:SD} \emph{(Synthetic data,~\cite{jordon2022synthetic})}
\textbf{Synthetic data} is data that has been generated using a purpose-built mathematical model or algorithm (the \textbf{generator}), with the aim of solving a (set of) data science task(s). 
\end{definition}

Let $D$ denote a tabular database describing data subjects through attribute set $\mathcal{A}(D)$. Rows $d\in D$ are $|\mathcal{A}(D)|$-tuples with a value $v(d,a)$ for each attribute $a\in \mathcal{A}(D)$. For $A=\left\{a_1,a_2,...,a_m\right\}\subseteq \mathcal{A}(D)$, we denote by $v(d,A)$ the tuple $(v(d, a_1), v(d, a_2),...,v(d, a_m))$. We denote by $\mathcal{D}$ the space of all possible rows over $\mathcal{A}(D)$.

In the remainder, we restrict our analysis to \emph{Synthetic data generated from original real-world data} (hereafter ``synthetic data''). We denote by $\hat{D}\sim G(D)$ that synthetic dataset $\hat{D}$ was obtained from a stochastic generator $G$ trained on an original dataset~$D$. This restriction rules out data that is generated through modeling processes that do not infer generation conditions directly from real-world data, such as simulation. Our form of synthetic data has become widespread in recent years, in part due to the success of machine learning-based generation methods~\cite{Figueira2022SurveySD, jabbar2020surveygGAN, Bellovin18, Exeter, jordon2022synthetic}. 

\subsection{Legal Background}
Synthesizing data from personal data inherently involves processing personal data, as the synthetic output is derived from an original dataset containing personal data. This principle parallels other anonymization processes. Consequently, such processing must fully comply with the relevant regulatory framework (e.g., the GDPR).
It is a separate question whether the synthetic data itself can be considered personal data or not. Depending on how effectively privacy risks have been mitigated during its generation, synthetic data may fall under either the personal or non-personal data category.

\subsubsection{Legal Definition of Personal Data.} GDPR Article 4(1), defines personal data as “any information relating to an identified or identifiable natural person (‘data subject’).” This definition comprises three essential elements: (i) “any information,” indicating the broad scope of data types that may be considered personal; (ii) “identified or identifiable natural person,” which includes both direct and indirect methods of identification; and (iii) “relating to,” which establishes the connection between the data and the individual \cite{WPpersonaldata, wp29anon}. The term “relating to” is particularly significant, as it can be interpreted in multiple ways—such as “in content” (where data explicitly describes an individual), “in purpose” (where data is processed to infer information about an individual), or “in result” (where the use of data indirectly affects an individual) \cite{WPpersonaldata, Cesar}.

\subsubsection{Anonymization and the law} Anonymity has long been understood as a decisive factor in determining the applicability of data protection law, including under the GDPR. According to the GDPR, an anonymization process should ensure that, once effective techniques have been applied, the data no longer carries the risk of identifying a specific individual and is thus exempt from GDPR’s scope. Recital 26 states that data should be considered anonymized when it has been “rendered in such a manner that the data subject is not or no longer identifiable.” Consequently, synthetic data should be regarded as anonymous only when it precludes any form of direct, indirect, or auxiliary identification. Yet, it would be misleading to interpret this as establishing a strict binary between personal and non-personal data. Rather, the notion of identifiability under EU law invites a more nuanced, context-aware reading—one that accounts for the risks and means of re-identification as they arise in specific technological and organizational settings.

\subsubsection{Reasonableness \& proportionality principles}\label{subs:RPP}
The principle of proportionality plays a central role in balancing the benefits of data processing with privacy protection. The possibility of re-identification—i.e., transforming anonymized data back into personal data—must be evaluated according to its likelihood within a given dataset. Recital 26 clarifies that determining whether an individual is identifiable requires considering all methods “reasonably likely” to be employed:

\begin{quote} \emph{To determine whether a natural person is identifiable, account should be taken of all the means \textbf{reasonably} likely to be used, such as singling out, either by the controller or by another person to identify the natural person directly or indirectly. To ascertain whether means are reasonably likely to be used to identify the natural person, account should be taken of all objective factors, such as the costs of and the amount of time required for identification, taking into consideration the \textbf{available technology at the time of the processing and technological developments}.} \end{quote}



In other words, the robustness of the anonymization process is gauged by the “means reasonably to be used” test \cite{WPpersonaldata}, and the risk of re-identification must be evaluated on the basis of what an adversary could realistically achieve, not in hypothetical or extreme scenarios. If identifying an individual requires unreasonable effort, time, or resources, the data can be considered anonymous and is therefore exempt from data protection rules. The feasibility of re-identification must always be assessed in relation to existing technologies and anticipated future developments.

\subsection{Anonymization benchmarks}\label{sec:WP29ao}
\subsubsection{WP29 typified risks}


As previously noted, data is considered anonymized when the data subject is "not or no longer identifiable," meaning that re-identification is rendered impossible. However, because this determination is inherently ex-post, assessing anonymization can be challenging. To facilitate these evaluations, the Article 29 Data Protection Working Party (WP29) established in 2014 three key risk benchmarks for effective anonymization: Singling Out, Linkability, and Inference \cite{wp29anon}. 

\begin{itemize}
    \item \textbf{Singling Out} corresponds to the possibility to isolate some or all records which identify an individual in the dataset;
    \item \textbf{Linkability} is the ability to link, at least, two records concerning the same data subject or a group of data subjects (either in the same database or in two different databases). If an attacker can establish (e.g. by means of correlation analysis) that two records are assigned to a same group of individuals but cannot single out individuals in this group, the technique provides resistance against ``singling out'' but not against linkability;
    \item \textbf{Inference} is the possibility to deduce, with significant probability, the value of an attribute from the values of a set of other attributes (exploiting statistical correlations within the data). In an attribute inference attack (AIA), if a dataset $D$ trains a generator $G(D)$ and synthetic data $\hat{D}\sim G(D)$ is accessible alongside auxiliary information, an adversary may combine these to infer unknown target attributes.
\end{itemize}

These legal risk benchmarks inform the adversarial threat models employed in our analysis. By mapping each WP29 risk to specific adversary objectives and capabilities, the threat model configurations developed in the following sections are designed to reflect what may be considered “reasonably likely” within the meaning of Recital 26 GDPR. In this way, our framework aligns technical risk assessment with the contextual and proportional nature of legal anonymization requirements.

Legally speaking, a robust anonymization procedure — that is, any data processing which produces output that, at a given time and within a specific context, can no longer be considered personal data — must effectively prevent these three typified re-identification risks. Building on these risk factors, this research proposes standardized frameworks for assessing synthetic data as a privacy-enhancing technology (PET) under the GDPR, thereby guiding data controllers, researchers, and policymakers.

\subsubsection{Alternative definitions of privacy risks}

The WP29 risks do not always align with technical classifications of typified privacy risks (or re-identification risks). Both Hu et al.~\cite{hu2023advancing} and Raab et al.~\cite{raab2024privacymetrics} classify risks into two categories: \begin{enumerate*}
    \item Identity disclosure: the ability to identify an individual in a dataset from a set of known characteristics;
    \item Attribute disclosure: the ability to find out a previously unknown attribute value associated with a record in the original data from a set of known characteristics.
\end{enumerate*}
These definitions align with those of ``singling out'' and ``inference'' under the WP29 to an extent. However, they differ from the WP29 definition in two key ways: \begin{enumerate*}
    \item Raab et al.'s~\cite{raab2024privacymetrics} definitions focus on \emph{what can be inferred} from an information release: either an individual's identity, or a value of one of their attributes. For this reason, linkability is not considered a separate risk;
    \item The role of auxiliary information (``known characteristics'') is directly incorporated into Raab et al.'s~\cite{raab2024privacymetrics} definitions of identity and attribute disclosure. This provides an idea of how an adversary would exploit a risk in an attack. The required \emph{auxiliary information} could be interpreted as linked data, technically making any form of attack an exploitation of linkability under WP29. 
\end{enumerate*}\\
\indent Besides WP29 and the definitions by Hu et al.~\cite{hu2023advancing} and Raab et al.~\cite{raab2024privacymetrics}, attacks compromising confidentiality from the field of adversarial machine learning~\cite{adversarial} are also commonly applied to synthetic data generators. In the context of synthetic data, the main risks (not due to deliberate system misuse) are: \begin{enumerate*}
    \item Membership inference attacks (MIAs): an attack in which the adversary can deduce whether a given record was in the training data of a given machine learning model, based on this model and/or its output. In the context of synthetic data, a MIA is successful if, for some record $d\in \mathcal{D}$, the attacker can infer whether or not $d\in D$ through access to $G(D)$ and/or $\hat{D}\sim G(D)$;
    \item Model stealing or shadow modeling: the adversary can reproduce the input-output relationship of a machine learning model (such as a synthetic data generator).
\end{enumerate*}\\
\indent Stadler et al.~\cite{groundhog} provide formalizations to model the WP29 attacks as specific forms of MIAs each. Both adversarial ML approaches tend to require considerable generator access, as introduced in Section~\ref{sec:threatmodel}.

\subsection{Threat models}\label{sec:threatmodel}
We assume that the adversary aims to obtain sensitive information about individuals whose data is contained in an original dataset used to train a synthetic data generator. Below, we describe adversary resources that can be exploited in privacy attacks when using synthetic data as a PET. These are the tools and information available to the adversary in conducting attacks. These resources can be subdivided into three categories: 
\begin{enumerate*}
    \item the synthetic dataset; 
    \item generator access; and 
    \item auxiliary information.    
\end{enumerate*} 

\indent Under {\bf no-box generator access}, the adversary exploits access to synthetic data without leveraging information about the model used to construct it. They may leverage auxiliary information sources, such as knowledge of particular individuals in the training data, including if this can be derived from the synthetic data's performance in downstream tasks; the source of the training data; modeling decisions and circumstances during generator development and training; the structure of the (synthetic) data (e.g. if attribute names are indicative of specific circumstances). This form of auxiliary information is formalized in Definition~\ref{def:auxil}.

\begin{definition}\label{def:auxil} \emph{(Auxiliary information)}
    Let $D$ be a dataset with attributes $\mathcal{A}(D)$. An adversary has \textbf{auxiliary information} if there exists a subset $A\subseteq \mathcal{A}(D)$ of attributes and a subset $D'\subseteq D$, such that the adversary knows the values $v(d,A)$ for all $d\in D'$. We refer to the attributes in set $A$ as {\bf quasi-identifiers}. We denote auxiliary information consisting of quasi-identifier set $A$ and records $D'\subseteq D$ by $\texttt{Aux}(A, D')$.
\end{definition}

\noindent {\bf Black-box generator access} is available when a synthetic data service provider offers a training algorithm for a generator to the end-user as a black box. This means the end-user can: \begin{enumerate*}
    \item Use the generator to produce an unbounded number of synthetic datasets;
    \item Train the generator on any available input dataset matching the algorithms capabilities.
\end{enumerate*}
However, the end-user does not have direct access to the model's internal routines and (hyper)parameters.

\indent {\bf White-box generator access} is available when the synthetic data provider provides not only access to the input-output relationship of their training algorithm (as under black-box access), but also to the full implementation of the algorithm. This could for instance be the scenario where a synthetic data provider makes their training algorithm open-source. It provides adversaries with full information on how the generator $G$ (a typically stochastic function) maps real data $D$ to synthetic data $\hat{D}\sim G(D)$. The conditions to carry out white-box or even black-box attacks are typically not met in practical scenarios~\cite{Emam2025}. \\


\section{Relation to prior research}\label{sec:prior}
Several works present \textbf{literature reviews and surveys} on synthetic data quality assessment, focusing on or including a classification of privacy evaluation methods. Kabaachi et al.~\cite{OtherSurvey} study utility and privacy quantification in a review of 92 papers. They find that, while 85\% (78/92) of papers include synthetic data for privacy use cases, only 42\% of papers (39/92) use a privacy quantification method. Of the papers conducting privacy quantification, 38\% (15/39) rely solely on either differential privacy or data masking for additional privacy protection. The remaining 62\% (24/39) employ at least one of a diverse set of privacy quantification methods.\\
\indent Hu et al.~\cite{hu2023advancing}, Osorio et al.~\cite{osorio2024privacy}, and Boudewijn et al.~\cite{Boudewijn} identify common privacy risks, classifying them into several categories. These include classifications based on potential risk factors (e.g. identity and attribute disclosure~\cite{hu2023advancing}), potential weaknesses of generated synthetic data (e.g. the use of skewness or background knowledge~\cite{osorio2024privacy}), and properties of data privacy quantification methods (e.g. mathematical properties, statistical indicators, attack simulations~\cite{Boudewijn}).\\
\indent\textbf{Experimental benchmarks for privacy quantification in synthetic data} include Hernandez et al.~\cite{Hernandez21} and Yan et al.~\cite{SurveyHealth}, who propose a number of synthetic data fidelity, utility, and privacy metrics and provide experimental benchmarks. Both works compute multiple metrics on multiple synthetic datasets, proposing the use of the metrics as benchmarks. However, these works primarily benchmark synthetic data as a PET, rather than the privacy quantification methods themselves. By contrast, our aim is to provide empirical benchmarks for privacy quantification methods. This could lead to a more nuanced understanding of their efficacy and sesnitivity to parameters and data properties. \\
\indent Two notable recent works involve \textbf{stakeholders or expert panels} in building consensus around synthetic data privacy quantification. Brunelli et al.~\cite{Shalini2024} present a synthetic data privacy and utility toolbox (\emph{``SURE''}, for \emph{``Synthetic Data: Utility, Regulatory compliance, and Ethical privacy''}) involving both statistical privacy indicators and attack simulations. The SURE-project directly invokes feedback of synthetic data end-users, namely data scientists and data protection officers (DPOs). A detailed feedback report of these users is still forthcoming. Pilgram et al.~\cite{Emam2025} adopt a similar approach to consensus building through expert feedback. Instead of direct stakeholders and end-users, they gather feedback through an expert panel consisting of long-term editorial board members of one journal and one conference on data privacy. They find that attack simulations are regarded positively, as they directly model a concrete risk (identity or attribute disclosure). The expert opinions in these studies are based on custom use of a specific toolbox and descriptions of metrics. Our work, by contrast, provides experimental results under several risk models, offering empirical insight into metrics' responses to controlled inserted risks.\\
\indent Legal scholars describe synthetic data as a promising privacy-enhancing technology (PET), particularly in contexts where access to granular data is necessary but privacy concerns are paramount~\cite{Bellovin18}. However, the application of the General Data Protection Regulation (GDPR) to synthetic data lacks clarity~\cite{Cesar}. The legal community appears aware of the need for quantitative privacy assessment, welcoming interdisciplinary research into standardization of privacy assessment in synthetic datasets, as the lack of consensus may hinder both innovation and regulatory compliance~\cite{Exeter}. Our work contributes to this research gap by showcasing the merits and weaknesses of prevailing privacy quantification methods theoretically and empirically.

\section{Synthetic Data Privacy Quantification Taxonomy}\label{sec:quant}
In this section, we propose a taxonomy of approaches to the quantification of the degree of privacy protection of synthetic datasets and their generators. The main distinction in our taxonomy centers around the practical approach to measuring privacy, categorized into the three approaches:
\begin{enumerate*}
    \item privacy properties;
    \item statistical privacy indicators; and
    \item attack simulations.
\end{enumerate*}

\begin{figure*}
\begin{tikzpicture}[level distance=1.25in,sibling distance=.25in,scale=.68]
\tikzset{edge from parent/.style= 
            {thick, draw,
                edge from parent fork right},every tree node/.style={draw,minimum width=1in,text width=1in, align=center},grow'=right}
\Tree 
    [. {Synthetic data privacy quantification}
        [.{Mathematical properties}
                [.{Generator properties}
                    [.{Differential Privacy} ]
                    [.{Other} ]
                ]
            [.{Data properties} 
                [.{$k$-Anonymity and related} ]
            ]          
        ] 
        [.{Statistical indicators}
                [.{Distance or similarity-based} 
                      [.{IMS} ]
                    [.{DCR} ]
                    [.{$k$-NN} ]
                ]
            [.{Other} ]
        ]
        [.{Attack simulations}
                [.{MIA} 
                    [.{No-box}
                        [.{MIAs with overfitting detection~\cite{vanbreugel2023DOMIAS, golob2024MamaMIA}} ]
                        [.{Uniqueness-based}~\cite{Anonymeter} ]
                        ]
                    [. {Black-box}
                        [.{Stadtler et al.~\cite{groundhog}} 
                        [.{MIA with vulnerable record discovery~\cite{meeus2023achilles}} ] 
                        ]
                        ]
                ]
            [.{Linkability} ]
            [.{AIA} 
                [.{No-box}
                    [.{Distance-based} ]
                    [.{Uniqueness-based} ]
                    [.{ML-based} ]
                ]
                [.{Black-box}
                    [.{Cf. black-box MIAs} ]
                ]
            ]
        ] 
    ]
\end{tikzpicture}\caption{Overview of synthetic data privacy quantification methods} \label{fig:tree}
\end{figure*}
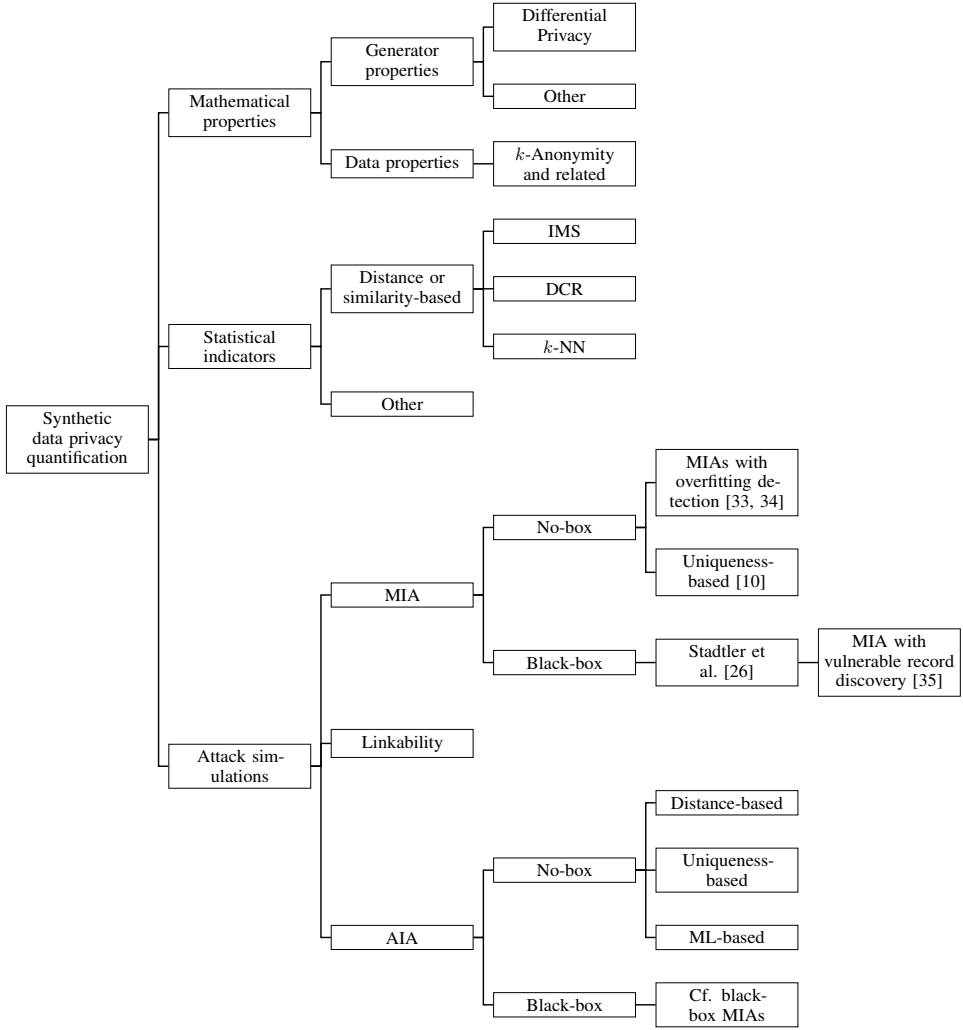

 Figure~\ref{fig:tree} provides an overview of the approaches to synthetic data privacy quantification. In the remainder of this section, we detail the elements in the taxonomy.

\subsection{Differential Privacy and other Privacy Properties}

\subsubsection{Differential privacy}
Differential privacy (DP, see~\cite{Dwork2014dif}), as formalized in Definition~\ref{def:DP}, is gaining traction as a privacy mechanism in synthetic data generation. 

\begin{definition}\label{def:DP}\emph{(Differential Privacy)}
A randomized algorithm $\mathcal{M}$ is $(\varepsilon,\delta)$-\textbf{differentially private} ($(\varepsilon,\delta)$-DP) if for any set $S$ of possible outputs of $\mathcal{M}$, we have:
\begin{equation}\label{eq:DP}
        \mathbb{P}\left[\mathcal{M}(D)\in S\right]\leq e^{\varepsilon}\cdot\mathbb{P}\left[\mathcal{M}(D')\in S\right]+\delta
\end{equation}
for all databases $D, D'$ such that $D'=D\setminus\left\{d\right\}$ for some $d$.  We refer to $(\varepsilon,0)$-DP simply as $\varepsilon$-DP.

\end{definition}


Intuitively put, DP limits the influence of any individual record $d$ in a given input dataset $D$ on the DP generator $G(D)$ through \textit{privacy budget} $\varepsilon$.  This makes it difficult for adversaries to target specific records in their attacks. The privacy budget is specified by the user, allowing them to directly control the degree of privacy introduced. This, along with mathematical guarantees (see, e.g. DP's immunity to post-processing~\cite{Dwork2014dif}) make DP a widespread privacy framework in practice.

A plethora of methods was introduced to obtain DP generators~\cite{DPGAN_Survey, DPGAN, yoon2018pategan, FL, NIST, ganev2023inadequacy}. DP is a property of the generator, added a priori. This is in contrast to metrics and attack simulations that are evaluated a posteriori, once the generator is fully trained and produces synthetic data. We therefore include DP in risk modeling, choosing a range of privacy budgets to control the amount of risk deliberately inserted for benchmarking (see Section~\ref{sec:riskmodel}). DP further allows datasets to be shared securely by responding to specific queries rather than releasing entire datasets, while enabling auditing of queries and protecting privacy through further anonymization techniques \cite{wp29anon}.

\subsubsection{$k$-Anonymity and related concepts}
While DP is a property of information release systems, $k$-anonymity and related properties pertain to anonymized datasets themselves. The underlying notion of $k$-anonymity is that \emph{uniqueness is a core risk factor}. For instance, while ``place of residence'' is not a direct identifier, publicizing a highly unique location can still put an individual at risk.\\
\indent A traditional approach to deal with such risks is \emph{generalization}. In the example, ``place of residence'' could be generalized to ``region''. This offers more privacy protection, as it makes it harder to single out individuals with unique locations. The concept of $k$-anonymity formalizes this form of protection. Formally, a dataset $D$ is \emph{$k$-anonymous} if any combination of quasi-identifier values present in $D$ occurs at least $k$ times. Generalization until $k$-anonymity is achieved divides the dataset up into equivalence classes with respect to the quasi-identifiers, each with a cardinality of at least $k$.

By itself, $k$-anonymity does not prevent AIAs~\cite{wp29anon}. Suppose in a $k$-anonymous set, city is generalized to country and age is generalized to age groups. However, 99\% of individuals in Italy aged 50-55 have a high income. Then an adversary cannot re-identify the record of a specific individual from Trieste, aged 52. However, they can infer that the individual is highly likely to have a high income. This problem of parameter inference is addressed by $l$-diversity, requiring that in each equivalence class, the sensitive attributes take on at least $l$ distinct values. The primary objective is to minimize equivalence classes with low attribute diversity, ensuring that attackers with prior knowledge about a specific individual face substantial uncertainty \cite{wp29anon}. While conceived as a privacy protection property, the concept of $l$-diversity can itself be exploited in AIAs. This is discussed in Section~\ref{sec:AIAquant}.

\subsection{Statistical Privacy Indicators}\label{sec:stats}
Statistical privacy indicators typically measure the occurrence of potentially hazardous synthetic records. In most cases, this corresponds to identifying if any synthetic records are more \emph{similar to specific original records} than statistically reasonable. In this paper, we the most common indicator categories, namely:
\begin{enumerate*}
    \item identical match share (IMS, also known as replicated uniques~\cite{Raab}, repU~\cite{raab2024privacymetrics}, and unique exact match~\cite{NIST});
    \item distance-to-the-closest-record (DCR); and
    \item $k$-nearest neighbor ($K$-NN)-based privacy metrics.    
\end{enumerate*}
Other statistical indicators for privacy assessment in synthetic data for instance include the maximum mean discrepancy (MMD) to test whether overfitting took place~\cite{smooth, esteban2017realvalued}, but are rare in practice.

\subsubsection{Identical match share} IMS metrics simply document how many of the synthetic records are also in the real dataset~\cite{NIST, Ebert, PanfIEEE}. In our experiments, we express this as a proportion if the number of rows in the real dataset, thus: $\texttt{IMS}(D, \hat{D}):=\frac{|D\cap\hat{D}|}{|\hat{D}|}$.

\subsubsection{Distance to the closest record} DCR metrics are computed by comparing two quantities: the \textit{Synthetic to Real distance} (SRD) and the \textit{real to real distance} (RRD). Let~$\texttt{Dist}:\mathcal{D}\times\mathcal{D}\rightarrow \mathbb{R}$ be a distance metric. For a given synthetic record $\hat{d}\in\hat{D}$, the  is the distance to its closest real record, as formalized in in equation~\eqref{eq:SRD}.

\begin{equation}\label{eq:SRD}
    \text{SRD}(\hat{d}) := \min_{d \in D} \text{Dist}(\hat{d}, d) 
\end{equation}

The RRD is typically computed using a \emph{holdout set}. Partition $D$ into two sets $D_1, D_2$. For a given record $d_1\in D_1$, the RRD is then the distance to its closest record in $D_2$. This is formalized in equation~\eqref{eq:RRD}.

\begin{equation}\label{eq:RRD}
    \text{RRD}(d_1) := \min_{d_2 \in D_2} \text{Dist}(d_1, d_2) 
\end{equation}

Suppose, for some synthetic record $\hat{d}\in\hat{D}$, we have a statistically unusually small SRD. This means that $\hat{d}$ is very close to some real record $d\in D$ in a manner that may not just be due to stochasticity. $\hat{d}$ may therefore leak sensitive information. DCR metrics assess whether such records exist by comparing the SRD and RRD distributions. In our experiments, we will use a percentile-based comparison, as is common in practice~\cite{Ebert, Platzer, PanfIEEE}. In particular, we compute the DCR as in equation~\eqref{eq:DCR}.

\begin{equation}\label{eq:DCR}
    \texttt{DCR}(\hat{D}, D) := \frac{\left|\left\{\hat{d}\in\hat{D}:\texttt{SRD}(\hat{d})<\texttt{RRD}_\alpha\right\}\right|}{\frac{\alpha}{100}\cdot|D_1|}
\end{equation}

where $\alpha$ is a percentage (we use 2\% throughout) and $\texttt{RRD}_\alpha$ is the $\alpha^{th}$ percentile of the RRD distribution. Equation~\eqref{eq:DCR} is the ratio of SRD values below the $\alpha^{th}$ RRD percentile and the number of RRD values that percentile (naturally $\alpha\%$ of $D_1$). 

\subsubsection{$k$-Nearest neighbors metrics} Indicators using $k$-nearest neighbors ($k$-NN) generalize DCR-based metrics by comparing neighborhoods of synthetic and real records. We define the synthetic to real $k$-neighborhood of a synthetic record $\hat{d}\in\hat{D}$, denoted by $N_{\texttt{SRD}}^k(\hat{d})$ as the $k$ nearest records to $\hat{d}$ in $\hat{D}$ with respect to the SRD. We define the real-to-real $k$-neighborhood $N_{\texttt{RRD}}^k(d)$ analogulously using the RRD. We then apply equation~\eqref{eq:DCR} with the means of the neighborhoods $N_{\texttt{SRD}}^k(\hat{d})$ and $N_{\texttt{RRD}}^k(d)$ in place of the SRD and RRD.

\subsection{Attack Simulation}\label{sec:attacks}
We include attack simulations tailored to the WP29 risks. All three of these risks can be implemented with different attack mechanisms. Below, we provide an overview of possible risk mechanisms and our implementations. 

\subsubsection{Membership Inference and Singling Out}
Identity disclosure is most closely modeled by membership inference attacks (MIAs)~\cite{Anonymeter, meeus2023achilles, vanbreugel2023DOMIAS}. In adversarial ML, a MIA is an attack in which the adversary can determine if a given individual's record was in the training dataset of a given model~\cite{adversarial}. In the context of synthetic data with no-box generator access, MIAs based on uniqueness~\cite{Anonymeter, frolich2024fraunhofer} and on distances~\cite{Goncalves, choi2018generating} are the most common.

{\bf Uniqueness-based MIAs} detect outliers in the synthetic data and propose these as potentially real original records. The underlying assumption is that outliers in the original data may be more susceptible to MIAs. By definition, they are located in a sparse region of the dataset, due to which the generator may locally overfit.

Anonymeter's singling out attack~\cite{Anonymeter} looks for unique records with respect to a subset of attributes in the synthetic dataset in a brute-force manner. Anonymeter models attacks as guesses. A singling out attack is essentially a guess of the form ``a record $d$ with parameter values $(v(d,a_1), v(d,a_2),...,v(d,a_n))$ is a unique record in the original dataset'', in which the combination of values is informed by the synthetic data. 

Let ${\bf g}:=\left\{g_1,g_2,...,g_{N_A}\right\}$ be a set of $N_A\in\mathbb{N}$ guesses and ${\bf o}=\left\{o_1,o_2,...,o_{N_A}\right\}$ their respective outcomes, where $o_i$, $i=1,2,...,N_A$ is defined through equations~\eqref{eq:o}.
\begin{equation}\label{eq:o}
o_i:=\begin{cases}
    1, &\text{if guess $g_i$ was correct}\\
    0, &\text{otherwise}
\end{cases}
\end{equation}
Guess effectiveness is interpreted as a Bernoulli trial with  success probability $\hat{r}$. Estimators $r, \delta_r$ can then be computed such that $\hat{r}\in r\pm\delta_r$ at confidence level~$\alpha$ via the Wilson Score Interval.\\
\indent We note that several novel approaches further explore the degree to which \emph{isolation} serves as an attack vector. Meeus et al.~\cite{meeus2023achilles} use a form of outlier detection to identify potential targets for black-box MIAs. D'Acquisto et al.~\cite{Acquisto} rigorously study the degree to which information release systems increase an adversary's abilities to isolate and subsequently identify real individual's information in a novel probabilistic manner.\\
\indent Domias~\cite{vanbreugel2023DOMIAS} is a MIA approach grounded in {\bf overfitting detection}. Unlike previous methods (e.g.,~\cite{hayes2018logan, chen2020gan, hilprecht2019monte}) which rely solely on synthetic data to infer membership, Domias explicitly targets overfitting by incorporating information from a holdout set of real records. The original dataset $D$ is partitioned into a training set $D_{\texttt{train}}$, a control set $D_{\texttt{control}}$, and a reference dataset $D_{\texttt{reference}}$. A generative model $G(\cdot)$ is trained on $D_{\texttt{train}}$ to produce a synthetic dataset $\hat{D}=G(D_{\texttt{train}})$. The core premise of this attack is that the generator overfits to outliers within the training set; consequently, these points exhibit a significantly higher density in the learned synthetic distribution than in the true underlying distribution. To quantify this, the method estimates the densities of the synthetic data and$D_{\texttt{reference}}$. The attack evaluates test points—drawn from $D_{\texttt{train}}$ (specifically targeting outliers) and $D_{\texttt{control}}$—using Eq.~\ref{eq:domias}:

\begin{equation}\label{eq:domias} 
    A(x^*) = f \left( \frac{p_{\texttt{synth}}(x^*)}{p_{\texttt{reference}}(x^*)} \right) 
\end{equation}

where $p_{\texttt{synth}}$ represents the learned distribution of the synthetic data, $p_{\texttt{reference}}$ is the learned distribution of the reference dataset, $x^*$ is the target point, and $f:\mathbb{R} \rightarrow[0,1]$ is a monotonically increasing function. If $x^*$ is an outlier from the training set, the numerator dominates the denominator. Conversely, if $x^*$ belongs to $D_{\texttt{control}}$, the numerator and denominator yield approximately equal values.


\indent \textbf{Black-box MIAs} generally operate as follows. For a given target record $d_t\in\mathcal{D}$, the adversary aims to predict whether $d_t$ was in the training set $D$ of synthetic dataset $\hat{D}\sim G(D)$. To do so, the adversary exploits their black-box access to the generator's training algorithm. This typically involves the following steps~\cite{groundhog, TAPAS, ganev2023inadequacy, meeus2023achilles, vanbreugel2023DOMIAS, golob2024MamaMIA}:
\begin{enumerate}
    \item The adversary takes $k$ datasets $D_1,...,D_k$ with $D_i\subset\mathcal{D}\setminus\left\{d_t\right\},~~i=1,...,k$;
    \item For each such dataset $D_i$, the adversary trains a generator $G(D_i)$ and produces an output dataset $\hat{D}_i\sim G(D_i)$. The adversary stores datasets $\hat{D}_i$ along with the label ``no target'';
    \item Next, the adversary creates the $k$ datasets $D_i':=D_i\cup {d_t}$ for $i=1,...,k$, obtained by adding the target record to the chosen datasets;
    \item The adversary now trains generators $G(D_i')$, producing synthetic datasets $\hat{D}_i'\sim G(D_i')$. The adversary stores the sets $\hat{D}_i'$ along with the label ``target'';
    \item The adversary now has a labeled collection of synthetic datasets. They leverage this collection to train a model $\mathcal{M}$ that predicts whether the target record $d_t$ was in the dataset used to train a generator $G$, based on an output synthetic dataset of $G$.    
\end{enumerate}
Step (5) usually involves feature extraction techniques to represent the collection of synthetic datasets as a table or other format of use to easily train $\mathcal{M}$ (typically a machine learning classifier). \\
\indent Given the high computational intensity of black-box attacks, methods that optimize target selection are highly preferable. Both vulnerable record discovery~\cite{meeus2023achilles} and strategies targeting overfitted regions~\cite{vanbreugel2023DOMIAS, golob2024MamaMIA} are favored specifically for their ability to significantly reduce this computational burden.".\\

\subsubsection{Linkability} To the best of our knowledge, Anonymeter's linkability attack~\cite{Anonymeter} is the only direct no-box implementation of linkability attacks. In this attack, the generator $G(D)$ is trained on original dataset $D$, including all of its attributes $\mathcal{A}(D)$, giving dataset $\hat{D}\sim G(D)$. Next, attribute set $\mathcal{A}(D)$ is partitioned into two (or more) mutually exclusive attribute sets $A_1$, $A_2$, such that $A_1\cup A_2=\mathcal{A}(D)$. The adversary's objective is now to link the values $v(d,A_1)$ and $v(d,A_2)$ together, giving the overall original record $d$, for $d\in D$.\\
\indent To do so, the adversary uses the synthetic dataset $\hat{D}$. In particular, they find the $k$ nearest synthetic neighbors (with respect to the SRD) to the partial records $\left\{v(d,A_1):d\in D\right\}$. Let $\mathbf{1}_{A_1}(d)$ be the set of indices of these $k$ nearest synthetic neighbors of $d\in D$. Then, they repeat this process for partial records $\left\{v(d,A_2):d\in D\right\}$, giving index sets $\mathbf{1}_{A_2}(d)$ of $k$ nearest synthetic neighbors of $d\in D$ with respect to attributes $A_2$. A linkability attack is now considered a success if $\mathbf{1}_{A_1}(d)\cap \mathbf{1}_{A_2}(d)\neq \varnothing$. In other words: a linkability attack against a record $d$ is successful if there is a $k^{\text{th}}$ synthetic nearest neighbor $\hat{d}$ of $d$ with respect to only attributes $A_1$ that is also a $k^{\text{th}}$ synthetic nearest neighbor of $d$ with respect to only attributes $A_2$. SRD calculations are based on the Gower distance. The same guess-based approach to attack efficacy is used as in the uniqueness-based MIA. Throughout, we use the same parameters as used in~\cite{Anonymeter}.

\subsubsection{Attribute inference}\label{sec:AIAquant}
Like identify disclosure attacks, attribute inference attacks (AIAs) can rely on several attack mechanisms. These again include uniqueness and distances, but also ML models trained to identify values of targeted parameters.

{\bf Uniqueness-based AIAs} seek to infer target $t$ attribute values by isolating unique (equivalence classes of) records in the synthetic dataset $\hat{D}$ with respect to auxiliary information $\texttt{Aux}(A,D')$. The Generalized Targeted Correct Attribution Probability (GTCAP,~\cite{chapelle2023TCAP}) does so by finding $l$-diverse equivalence classes in $\hat{D}$ with respect to $t$ with a low $l$-value. Suppose a dataset is anonymized through generalization until $k$-anonymity is achieved for non-target attributes. Then, the set has several equivalence classes with at least $k$ records each, for which all non-target attributes are the same (or in the same radius for continuous attributes). But if the $l$-diversity is low in a given equivalence class, then all these records will have similar values of target attribute $t$. 

The $l$-diversity with low $l$ of equivalence classes with respect to a specific sensitive attribute makes the class susceptible to a parameter inference attack. GTCAP is an approach to probabilistically quantify this susceptibility.  The GTCAP is a generalization of the TCAP~\cite{Elliott2015TCAP}, defined for synthetic record $\hat{d}\in\hat{D}$ by equation~\eqref{eq:TCAP}, where the square brackets are Iverson brackets and $A\subseteq \mathcal{A}(D)$ is a set of quasi-identifiers in the auxiliary information of the adversary.

\begin{equation}\label{eq:TCAP}
    \texttt{TCAP}(D,\hat{d}):=\frac{\sum_{d\in D} \left[v(d,A)=v(\hat{d},A),~ v(d,t)=v(\hat{d},t)\right]}{\sum_{d\in D} \left[v(d,A)=v(\hat{d},A)\right]}
\end{equation}

The GTCAP generalizes the TCAP for continuous attributes, by replacing equality with membership in spheres of given radius around $v(\hat{d},A)$ and $v(\hat{d},t)$.

In {\bf distance-based AIAs} (also known as majority-voting attack; local neighborhood attack, see, e.g.,~\cite{Anonymeter, TAPAS, Goncalves, choi2018generating}), adversaries exploit similarities between synthetic records and records from their auxiliary information (see Definition~\ref{def:auxil}). Suppose an adversary has access to: 
\begin{enumerate*}
    \item a synthetic dataset $\hat{D}$; and
    \item auxiliary information $\texttt{Aux}(A,D')$.    
\end{enumerate*}
In a distance-based attack, the adversary simply estimates the value $v(d',t)$ of record $d'\in D'$ and target attribute $t\in \mathcal{A}(D)\setminus A$ by assigning it the value of the nearest synthetic record, i.e.: $v(\hat{d}^*,t)$ for $\hat{d}^*=\argmin_{d \in D} \text{Dist}(\hat{d}, d')$. In some cases, this is generalized to an aggregate of the $k$ nearest synthetic neighbors of $d'$. However, experiments typically show that $k=1$ leads to the best attack performance~\cite{Goncalves, chen2019TCAP}.

{\bf ML-based AIAs} (also known as ``inference-on-synthetic''~\cite{TAPAS}) operate in a similar manner to distance-based AIAs, but instead of relying on distances, they rely on ML models. The adversary again has access to synthetic dataset $\hat{D}$ and auxiliary information $\texttt{Aux}(A,D')$. They want to use $\hat{D}$ to estimate the values of a target attribute $t\in \mathcal{A}(D)\setminus A$ for the records in $D'$. To do so, they train an ML model $M$, such that $M(v(\hat{d},A))\approx v(\hat{d},t)$ for $\hat{d}\in\hat{D}$. They then apply $M$ to the auxiliary information to estimate the target attributes of real individuals' records $v(d,t)$. Model $M$ is a classifier in case $t$ is categorical and a regression model in case $t$ is numeric.

{\bf Black-box AIAs} can be conducted as a sequence of black-box MIAs~\cite{groundhog}. Suppose that $d\in D$ is a target record, for which auxiliary information (in the sense of Definition~\ref{def:auxil}) is available for attributes $A\subseteq\mathcal{A}(D)$. Suppose now that $t$ is a categorical target attribute with range $T$. An adversary can then conduct $|T|$ MIAs, one for each possible value of $t$. If $d$ was in the training data, then the MIA with $d$'s value of $t$ will be successful.

\section{Empirical Assessment Framework for Tabular Synthetic Data Privacy Metrics}\label{sec:riskmodel}
To empirically assess privacy quantification methods' efficacy, we introduce the concept of a ``\emph{risk model},'' i.e. \emph{a manner to deliberately introduce privacy risks in synthetic datasets with the purpose of assessing how well privacy quantification methods identify these inserted risks}. Below, we introduce three risk models studied in our experiments, motivating their use and detailing their specific parametrization. 

\subsection{Risk models}
\subsubsection{Leaky risk model} 
By leaky risk model, we refer to the following experimental setup. The real dataset $D$ is divided into three sets: training ($D_t$), control ($D_c$), and release ($D_r$). The ``synthetic'' dataset $\hat{D}$ is then obtained by taking $D_r$ and adding a fraction $f_l:=\frac{|\hat{D}\cap D_t|}{|D_t|}$ of records from the training set. The leaky risk model allows us to control the direct impact of leaked private information on the privacy quantification method. By inserting risks in controlled unit steps, we can observe whether risk metrics respond to inserted risk linearly. Ultimately leaking the entire original dataset ($f_l=1$) allows us to observe whether the risk measures reach their theoretical maximum when all original records are leaked. 

\subsubsection{Overfitting risk model} 
The leaky risk model provides insight into the response of the assessment frameworks to controlled, deliberately inserted privacy leaks. However, this is not how privacy risks emerge in practice when using synthetic data as a PET. In practice, generator overfitting is considered a main source of risk~\cite{carlini2022outliers, groundhog, Boudewijn}. When a generator overfits, it might memorize specific patterns from the real data, rather then representing them stochastically. Thereby, they can put involved real individuals at risk by inferring and reproducing their personal information.

To account for risks due to generator overfitting, we proceed as follows. We measure models' performance through validation loss (see Appendix~\ref{app:vallos}). First, we train the generator optimally, giving optimal loss function value $L^*$. We define the \emph{overfit ratio} $f_o$. For a given overfit ratio $f_o$, we then retrain the generative model, but terminate training prematurely once its loss function satisfies $L=f_o\cdot L^*$. This allows us to evaluate the risks in the generator's output synthetic data when it is trained to a factor $f_o$ of optimality.\\
\indent When $f_o=1$, we have $L=L^*$, so no deliberate overfitting took place. When $f_o=2$, the generator's loss function value is twice that of its previously determined optimum. 

\subsubsection{Differential Privacy}\label{sec:DPriskmodel}
A third approach to deliberate risk insertion is through DP. In particular, we vary the privacy budget $\varepsilon$ of $\varepsilon$-DP. This allows us to assess how the privacy quantification methods to DP as a privacy mechanism. We then evaluate the metrics on their output synthetic datasets. By using large $\epsilon$ values (up to 100), we aim to simulate conditions where the differential privacy mechanism is nearly disabled, effectively mimicking an overfitting scenario. This allows us to illustrate how model behavior changes as privacy constraints are relaxed.

\subsection{Assessing robustness of privacy quantification methods}
\subsubsection{Outliers}
Privacy quantification methods may be sensitive to their parametrization, or to specific data properties, e.g. the presence of outliers~\cite{groundhog,carlini2022outliers,MIAnonOutlier}. We therefore propose combining the threat models with systematic outlier removal to assess these vulnerabilities. In particular, we propose repeating the experiments after removing $x$\% of outliers from the original datasets, with $x$ ranging from $0\%$ up to $10\%$. Generators should then be retrained on the outlier-removed original sets, after which the experiments are repeated. In our experiments, we let $x$ take values of $1$\%, $2$\%, $5$\%, and $10$\%, showcasing the relation between the number of removed outliers and privacy. We conducted these experiments for overfit ratios of both $f_o=1$ and $f_o=1.6$ to compound the effects of outlier removal and deliberate overfitting.

\subsubsection{Privacy quantification methods' parametrization}
The quantification methods may also be sensitive to specific parameters. In our experiments, we account for this through the following additional experiments:
\begin{enumerate*}
    \item we assess $k$-NN-based privacy indicators with different values of $k$;
    \item for the GTCAP, we repeat the experiments with several radii;
\end{enumerate*}
Some of the obtained findings seem general in nature and are corroborated by prior work, such as $k$-NN metrics being optimal with $k=1$~\cite{choi2018generating, Goncalves}. For GTCAP, practitioners should make context-sensitive decisions regarding the importance of parameters.

\subsection{Accounting for specific versus general inference}\label{sec:baseline}
\subsubsection{On ``Relating to'' in the GDPR}
Understanding what it means for data to ``relate to'' an individual is central to determining whether data counts as personal under the GDPR and whether synthetic data qualifies as anonymized. The legal literature, including Cesar~\cite{Cesar}, draws a threefold distinction: data may relate to an individual \textit{in content}, \textit{in purpose}, or \textit{in result}. Each of these has different implications for assessing privacy risks.

\textbf{``Relating to in content''} refers to data that contains specific information about an identifiable individual---for example, ``Jan Jansen is poor.'' This form is the most direct and is widely recognized as personal data under the GDPR.

\textbf{``Relating to in purpose''} applies when data, though not directly identifying, is  with the intention of affecting, evaluating, or making decisions about an individual. For instance, consider synthetic data used to generate a behavioral profile for targeted advertising. Even if the underlying data is anonymized, if the synthetic model is used with the purpose of tailoring content to ``a likely diabetic user in postcode 1234''---and this targeting impacts Jan Jansen---it may still count as relating to him ``in purpose.'' Here, the relevant behavior being influenced is not necessarily the data subject’s own, but rather the behavior of those interacting with or making decisions about the data subject.

\textbf{``Relating to in result''} refers to outcomes that affect individuals based on insights derived from data. For example, a generator trained on real health data may not contain Jan Jansen’s records, but if its predictions lead to changes in insurance policy pricing that impact him specifically, it arguably ``relates to'' him in result.

These nuances matter because only ``in content'' data necessarily involves information \emph{specific to an identifiable person}. ``In purpose'' and ``in result'' cases often deal with general, population-level insights. Treating these general inferences as personal data risks overextending the GDPR’s scope, and may conflate statistical accuracy with privacy breaches. A synthetic dataset that closely matches population-level statistics is not inherently privacy-violating. However, if a generative model overfits and memorizes information that is highly specific and unique to an individual, then the synthetic output may include traces of ``in content'' data, amounting to a genuine privacy leak.

To operationalize this distinction, we adopt and build upon baseline approaches from Giomi et al.~\cite{Anonymeter} and Sundaram et al.~\cite{annamalai2024linear}, which allow us to differentiate between information deducible from general patterns and information that can only result from direct memorization. 

\subsubsection{Anonymeter control set baseline}\label{subs:Anonymeterbase}
Giomi et al.~\cite{Anonymeter} propose the use of control sets in attack simulations. This works as follows. Subdivide the original dataset $D$ into a training set $D_{\texttt{train}}$ and a control set $D_{\texttt{control}}$. Train the generator on $D_{\texttt{train}}$ only. Evaluate the efficacy of the attacks on $D_{\texttt{train}}$, giving an effectiveness metric value $e_{\texttt{train}}$. Next, apply the the same attack to the $D_{\texttt{control}}$, giving effectiveness score $e_{\texttt{control}}$. This score indicates how well the attack performs against data that was not to train $G(D_{\texttt{train}})$. As such, efficacy is merely due to the \emph{general} utility of $D$, not to memorization of $G(D_{\texttt{train}})$. 

Next, compute the proportion $E$ to which the efficacy of the attack on $D_{\texttt{train}}$ is a consequence of information gained through $G(D_{\texttt{train}})$'s possible memorization of information in $D_{\texttt{train}}$. This is computed using equation~\eqref{eq:baseline}, where $e^*$ is the worst-case behavior of $e_{\texttt{train}}$, that is: the value of $e_{\texttt{train}}$ in case all attacks on $D_{\texttt{train}}$ are successful.

\begin{equation}\label{eq:baseline}
    E := \frac{e_{\texttt{train}} - e_{\texttt{control}}}{e^* - e_{\texttt{control}}}
\end{equation}

Attack effectiveness on the control set ($e_{\texttt{control}}$) is not due to memorization of specific information, but rather due to the degree to which the information of the synthetic data generalizes to its entire underlying population. The use of the baseline thus accounts for the distinction between specific training data and general population information.

\subsubsection{Canary records baseline for AIAs}\label{sec:canary} Houssieau et al.~\cite{TAPAS} and Sundaram et al.~\cite{annamalai2024linear} note that the success of an AIA may be due to general correlations between attributes in the data (i.e. general patterns), rather than specific learned information about individual records. This is referred to as the ``\textbf{base-rate problem}''. An individual's wealth may, for instance, correlate strongly with their education level, age, and place of residence. An AIA can then succeed by exploiting this \emph{general}, population-level correlation, rather than by exploiting specific information memorized or leaked about the individual (``\emph{in content}'') by the synthetic dataset or generator.\\
\indent Sundaram et al.~\cite{annamalai2024linear} propose accounting for the degree to which AIAs' efficacy is due to general correlations rather than memorization as follows. Let $D$ be a real dataset and let $t$ be a target attribute. For target record $d$, let $d'$ be defined through equation~\eqref{eq:canary}.
\begin{equation}\label{eq:canary}
  v(d',a) = \begin{cases}
    v(d,a), &\text{if } a\neq t\\
    r, &\text{if } a=t
\end{cases}  
\end{equation}

where $r$ is a random value sampled from the uniform distribution over the range of  $t$. Note that this transformation is applied deliberately to the target attribute of each target record. General correlations between $t$ and remaining attributes then have no impact on $d'$. Let $D'$ be the dataset obtained by taking $D$ and replacing $d$ with $d'$. Suppose we use $\hat{D}'\sim G(D')$ in an AIA with target record $d'$ and target attribute $t$. If this AIA is successful, this cannot be due to general correlations, due to the randomization. Thus, if the AIA is still successful, this indicates that the generator may leak \emph{specific} information relating to the individual corresponding to record $d$ ``\emph{in content}''. In our experiments, we conduct additional experiments with canary record baselines by randomly selecting $100$ canary records per dataset and averaging the success rates of AIAs against these canaries.

\section{Experiments}
\subsection{Synthetic data generation}

The experiment with synthetic datasets was conducted using the following methods: RealTabFormer~\cite{solatorio}, PATEGAN~\cite{yoon2018pategan}, Synthpop~\cite{synthpop}, AIM~\cite{AIM}. These models were selected to ensure a diverse representation of synthetic data generation approaches. They span a range of methodologies, including classical statistical techniques (Synthpop), deep generative adversarial networks (PATEGAN), and transformer-based architectures (RealTabFormer). In particular, this set includes both differentially private (PATEGAN, AIM) and non-differentially private (Synthpop, RealTabFormer) models, allowing for a comparative evaluation of privacy metrics across fundamentally distinct privacy-preserving paradigms. By selecting models that differ in architecture, training strategies, and privacy mechanisms, our evaluation aims to provide insights that are not limited to a single class of generative models but rather reflect broader trends and trade-offs in the field of synthetic tabular data generation.

RealTabFormer is a transformer-based framework designed for generating non-relational tabular data using an autoregressive model. Additionally, it can generate relational data by leveraging a sequence-to-sequence (Seq2Seq) architecture. In our experiments, we focus solely on the autoregressive model, specifically employing the GPT-2 architecture.\\
\indent PATEGAN is a generative adversarial network designed for generating synthetic tabular data that uses the PATE mechanism~\cite{papernot2016semi}. This mechanism incorporates differential privacy, a formal privacy guarantee that limits the amount of information about individual records in the training data that can be inferred from the output. It ensures that synthetic data generated respects privacy constraints, making it safer to use for sharing or analysis in sensitive domains like healthcare or finance.
\indent Synthpop is an R package that generates synthetic tabular data using a sequential modeling approach. It synthesizes variables one at a time, modeling the conditional distribution of each variable based on the variables already synthesized. It uses statistical modeling models like decision trees or regression.\\
\indent AIM is a differentially private model that follows the select-measure-generate paradigm. This approach involves three key steps:
\begin{enumerate*}
    \item Selection: Choosing queries, typically low-dimensional marginals, from a collection;
    \item Measurement: Adding noise to the selected queries to ensure differential privacy; and
    \item Query Refinement: Retaining the queries that best explain the noisy measurements.
\end{enumerate*}
The AIM framework iteratively performs the selection and measurement steps to progressively enhance the quality of the selected queries.

\subsection{Metrics evaluation}
Below, we detail the specific implementations of privacy quantification methods outlined in Section~\ref{sec:quant}. We restrict the experiments to the metrics most common in practice, namely no-box models. These prevail in \begin{enumerate*}
    \item {\bf academia}: see the literature reviews discussed in Section~\ref{sec:prior};
    \item {\bf industry}: Ganev et al.~\cite{ganev2023inadequacy} outline metrics of twelve synthetic data providers, while~\cite{ClearBox, frolich2024fraunhofer, avatar} describe metrics of one additional providers each. All rely on no-box metrics;
    \item {\bf Governmental organizations} such as the European Union \emph{Joint Research Centre (JRC)} and the United States \emph{National Institute of Standards and Technology} use no-box statistical privacy metrics for synthetic data privacy assessment~\cite{Ebert, NIST}.
\end{enumerate*}   
This restriction aligns with the reasonableness and proportionality principle, as more advanced threat models typically become computationally intensive. Furthermore, the synthetic data provider has direct control over generator access. More advanced threat models also often require highly specific assumptions. For instance, DOMIAS~\cite{vanbreugel2023DOMIAS} and MAMA-MIA~\cite{golob2024MamaMIA} require that the generator is marginals, conditionals, or copulas-based. Table~\ref{tab:commands} provides an overview of the metrics compared in the experiments.\\
\indent All metrics are evaluated using the risk models from Section~\ref{sec:riskmodel}. For the leaky risk model, we range the leak fraction $f_l$ from zero (no leaks) to one (all training data leaked) in equal steps of $0.2$; for the overfitting model, we range the overfit ratio $f_o$ from one (no deliberate overfitting) to two ($L = 2L^*$) in equal steps of $0.2$. For the DP risk model, we use the OpenDP Smartnoise AIM~\cite{AIM} and PATEGAN~\cite{yoon2018pategan} model to train $\varepsilon$-DP synthetic data generators with privacy budget $\varepsilon$ in \{$1.0$, $5.0$, $10.0$, $50.0$, $100.0$\}.

\begin{sidewaystable*}
\footnotesize
  \caption{Synthetic data degree of privacy quantification methods used in this study. Risk: risk measured or controlled; Aux: auxiliary information; MIA: membership inference attack; AIA: attribute inference attack; disc.: disclosure; SO: singling out; Link: linkability; ML: machine learning; IoS: inference-on-synthetic; LN: local neighborhood. Exactly the attribute disclosure attacks require access to auxiliary information. Nb: examples are for reference only: they may implement same attack methods and mechanisms in different manner than our implementations.}
  \label{tab:commands}
  \begin{tabular}{ccccccc}
    \toprule
    &&&&&& {\bf Our} \\ 
    {\bf Method}               &  {\bf Type}              &  {\bf Risk}   &  {\bf Mechanism} & {\bf Aux.} & {\bf Example(s)} & {\bf implementation}\\\midrule
    Differential privacy & Generator property    & Propensity             & NA   & No     & \cite{AIM, DPGAN_Survey} & OpenDP AIM \\ 
    Differential privacy & Generator property    & Propensity             & NA   & No     & \cite{yoon2018pategan} & Synthcity PATEGAN~\cite{https://doi.org/10.48550/arxiv.2301.07573} \\\midrule
    IMS                  & Statistical indicator & Propensity          & NA      & No  & \cite{Raab, NIST, Ebert, PanfIEEE} & Standard \\
    DCR                  & Statistical indicator & Propensity          & NA      & No  & \cite{Ebert, PanfIEEE}& Percentiles-based\\
    $k$-NN               & Statistical indicator & Propensity          & NA      & No  & \cite{Ebert, PanfIEEE}& Percentiles-based\\\midrule
    Outlier-based MIA    & Attack                & SO & uniqueness & No & \cite{Anonymeter} (SO), \cite{meeus2023achilles} & Anonymeter~\cite{Anonymeter} SO\\
    Outlier-based MIA (DOMIAS)   & Attack                & SO & uniqueness & Yes & \cite{vanbreugel2023DOMIAS} & ROC AUC classifier\\\midrule
    Linkability attack   & Attack                & Link & distance & Yes & \cite{Anonymeter} (Link) & Anonymeter~\cite{Anonymeter} Link\\\midrule
    Classifier inference & Attack                & AIA & ML       & Yes &\cite{TAPAS} (IoS)& XGBoost, accuracy \\
    Regression inference & Attack                & AIA & ML       & Yes & & XGBoost, RMSE \\
    Distance-based AIA& Attack            & AIA & distance & Yes & \cite{Anonymeter} (AIA), \cite{TAPAS} (LN), \cite{Goncalves, choi2018generating}  & Anonymeter~\cite{Anonymeter} AIA\\
    GTCAP                & Attack                & AIA & uniqueness & Yes & \cite{chapelle2023TCAP, Elliott2015TCAP, Taub2019TCAP, chen2019TCAP} & See~\cite{chapelle2023TCAP}\\

    \bottomrule
  \end{tabular}
\end{sidewaystable*}

\subsubsection{Statistical indicators} Implementation of the IMS is straight-forward. For DCR and $k$-NN, we use 2\% percentiles. We evaluate the Euclidean distance in an embedded space obtained through the embedding methods from~\cite{PanfIEEE}. To ensure that the measure yields a value of zero when no information is leaked and a value of one when the entire data is leaked, we normalize the DCR between its best case and its worst case. Thus, a privacy score of zero indicates no perceived risks, and a value of one indicates that all synthetic records pose risk to real records. 

\subsubsection{MIAs} for the uniqueness-based MIA, we use Anonymeter's singling out attack~\cite{Anonymeter}. We conduct experiments with outliers with respect to single attributes, as well as experiments in with the number of attributes varying between three and twelve, each consisting of at most\footnote{The number of guesses is dependent on how many predicates single out unique synthetic records. This may be fewer than 2000 if no more vulnerable records can be detected.} 2000 attacks ($N_A=2000$). The final reported risk is the highest risk $R$. Bootstrapping with $n=1000$ is used to measure the variance of the methods.\\
\indent For MIAs with overfitting detection, we adopt the approach introduced by van Breugel et al.~\cite{vanbreugel2023DOMIAS}. This method involves fitting two instances of a chosen density estimator (such as the Kernel Density Estimator (KDE) or the Block Neural Autoregressive Flow (BNAF)~\cite{de2020block}): one on the reference set and another on the synthetic dataset. For each test point, densities are computed using these fitted estimators. The DOMIAS score is then calculated as the ratio between the synthetic density and the reference density. Finally, the area under the ROC curve (AUC-ROC) is computed to evaluate the attack. The resulting AUC values were normalized to restrict the measure to the interval $[0,1]$, allowing for a standardized interpretation. The results, presented in Appendix~\ref{app:full}, demonstrate that the presence of outliers in the original data is crucial for this attack's success. The attack's mechanism hinges on the synthetic dataset being overly similar to the real one. When this happens, the synthetic probability density function (PDF) essentially memorizes the training data's distribution. This memorization, in turn, causes the synthetic PDF to assign high density values to the original training records, while assigning only medium values to records from other sets. Our findings perfectly align with this principle. For instance, the leaky and overfitting setups showed an increasing trend in attack success, which is expected as both scenarios produce synthetic data that is too faithful to the training set. On the other side, the attack's performance degraded in the outlier removal experiment. This confirms our hypothesis, as the very records that contribute most to the identifiable signal (the outliers) had been removed from the learning set, making the training data much harder for the synthetic PDF to distinguish.

\subsubsection{AIAs} In our experiments, numerical data was normalized with a MinMaxScaler. Hence, we consistently chose of the radii in the interval $(0, 1)$. In particular, we apply the GTCAP radii of 0.1. We use use $A=\mathcal{A}(D)\setminus\left\{t\right\}$ to test the worst-case scenario in which the adversary has access to all but one attribute of a given record.
\indent To assess distance-based AIAs, we use the the Anonymeter inference attack~\cite{Anonymeter} as an instance of distance-based AIAs. Like the singling out attack, this is evaluated in the form of an estimator of guess effectiveness. In our experiments, we use the same parameters as in~\cite{Anonymeter}.\\
\indent To conduct ML-based AIAs, we used the default unregularized \emph{Scikit Learn} implementation of eXtreme Gradient Boosting (XGBoost). For the Adult dataset, we used a classifier, evaluated with accuracy. For the Census and Texas datasets, we used a regressor evaluated using the root mean square error (RMSE). For classification models, we evaluate the accuracy. For regression models, we evaluate the root mean square error (RMSE). To bring RMSE values into a range comparable with accuracy metrics, the normalization shown in equation (\ref{eq:nrmse}) is applied: the original RMSE is divided by the data range, resulting in values within the interval $[0, +\infty]$, where 0 represents perfect model performance and 1 indicates that the prediction error is comparable to the variability of the original data. Values exceeding 1 suggest poor model performance. An additional normalization factor is introduced to shift the range to $[-\infty, 1]$, where 1 corresponds to perfect model performance.
\begin{equation}\label{eq:nrmse}
    \text{NRMSE}(y, \hat{y}) = 1 - \frac{\text{RMSE}(y, \hat{y})}{\text{range}(y)}
\end{equation}
Recall, this heuristic is designed to align accuracy metrics for categorical data and regression metrics for numerical data within a similar range of values, where a perfect model is consistently represented by a value of 1 across both types of metrics. However, it is important to recognize the underlying differences between the metrics and avoid overinterpreting normalized values as directly equivalent to accuracy.


For Linkability, we follow use the default Anonymeter implementation~\cite{Anonymeter}. 

\subsection{Datasets and experimental set-up}
We tested the methodology on three commonly used public datasets, namely: Adult~\cite{misc_adult_2}, the Texas Inpatient Public Use Data File (``Texas'') \cite{misc_texas}, and the 1940 Census full enumeration from IPUMS USA (``Census'')~\cite{misc_census}. The adult dataset has 48842 rows and fifteen attributes, of which six are numerical (float), and the remaining nine are categorical. The Texas and Census datasets are considerably larger, at respectively 193 and 97 attributes. We limit our experiments to subsets  of 28 and 37 attributes for these sets, respectively. We  take randomly selected, mutually exclusive subsets of cardinality 25,000 (Census) and 20,000 (Texas) as training and control sets for the Texas and Census data. See Appendix~\ref{app:data} for more information. All experiments were conducted on a computer equipped with AMD Ryzen Threadripper 2950X 16-Core Processor (CPU) and 128 Gib of RAM.

\subsection{Results}
Table~\ref{tab:res} contains the results with no inserted risk and maximal inserted risk for each of the metrics, dataset, and risk model. Appendix~\ref{app:full} shows the full results, as well as the correlation matrices of the privacy quantification methods per experiment for all risk models. Note that while each risk measure is normalized, different methods measure different quantities. Comparisons should therefore be based on the \emph{response to inserted risk}, rather than the exact values of the measurements. Table~\ref{tab:resTime} contains the computation times of the metrics per dataset and risk model.

\begin{sidewaystable*}
  \caption{Results of the leaky and overfitting risk models. RTF: RealTabFormer; O: outlier; D: distance; ML: machine learning. By ``no risk'', we indicate that no risk was deliberately added, i.e. $f_l=0$ for the leaky risk model; $f_o=1$ for the overfitting risk model; for the DP risk model, we equate ``no risk'' to a privacy budget of $\varepsilon = 0$. Simliarly, ``max risk'' refers to $f_l=1$; $f_o=2$; and $\varepsilon=100$. We use the asterisk (*) to denote the maximum risk, which exceeds any risk achieved with the previous values of $f_l$, $f_o$, or $\varepsilon$.}
  \label{tab:res}
  
  \resizebox{\textwidth}{!}{%
  \begin{tabular}{c|ccc|ccc|ccc|ccc|ccc}
    \toprule
                         & \multicolumn{3}{c|}{{\bf Leaky}}                & \multicolumn{6}{c|}{{\bf Overfit}}  & \multicolumn{6}{c}{{\bf DP}}             \\\midrule
                                        &                &         &      & \multicolumn{3}{c|}{{\bf RTF}} & \multicolumn{3}{c|}{{\bf Synthpop}} & \multicolumn{3}{c|}{{\bf PATEGAN}} & \multicolumn{3}{c}{{\bf AIM}} \\
{\bf Method}         &  {\bf Adult}  &  {\bf Texas}   &  {\bf Census} &  {\bf Adult}  &  {\bf Texas}  &  {\bf Census} &  {\bf Adult}  &  {\bf Texas}  &  {\bf Census} &  {\bf Adult}  &  {\bf Texas}  &  {\bf Census}  &  {\bf Adult}  &  {\bf Texas}  &  {\bf Census}\\
    \midrule
    IMS (no risk)    & 0.0 & 0.0 & 0.0 & 0.0    & 0.0    & 0.0010 & 0.0 & 0.0 & 0.0037  & 0.0 & 0.0 & 0.0 & 0.0 & 0.0 & 0.0 \\
    IMS (max risk)   & 1.0 & 1.0 & 1.0 & 0.8684 & 0.0162 & 0.9428 & 0.4377 & 0.0 & 0.1519 & 0.0 & 0.0 & 0.0 & 0.0 & 0.0 & 0.0\\
    IMS stdev (max risk) & 0.0045 & 0.0038 & 0.0034 & 0.0034 & 0.0008 & 0.0017 & 0.039 & 0.0 & 0.0015 & 0.0 & 0.0 & 0.0 & 0.0 & 0.0 & 0.0 \\
   \midrule
    DCR (no risk)    & 0.0011 & 0.0082 & -0.0013 & 0.0008 & -0.0009 & 0.0046 & 0.0002 & -0.0203 & 0.0064 &  -0.0204 & -0.0204 & -0.0204 & -0.0057 & -0.0204 & -0.0186 \\
    DCR (max risk)   & 1.0    & 1.0    & 1.0     & 0.5325 & 0.4783  & 0.5700 & 0.2933 & -0.0204 & 0.1115 & -0.0204 & -0.0204 & -0.0204 & 0.0019 & -0.0201 & -0.0046\\
    DCR stdev (max risk) & 0.0001 & 0.0 & 0.0 & 0.0011 & 0.0010 & 0.0010 & 0.0032 & 0.0 & 0.0005 & 0.0 & 0.0 & 0.0 & 0.0008 & 0.0001 & 0.0005\\
   \midrule 
    MIA, O (no risk)    & 0.0060 & 0.0400 & 0.0101 & 0.0278 & 0.0310 & 0.0270 & 0.0148 & 0.0153  & 0.0285 & 0.0136 & 0.0213 & 0.0010 & 0.0146 & 0.0072 & 0.0137 \\
    MIA, O (max risk)   & 0.9990 & 0.9990 & 0.9989 & 0.7620 & 0.6744 & 0.7895 & 0.4836 & 0.0257 & 0.1025 &  0.0101 & 0.0239 & 0.0010 & 0.0627 & 0.1279 & 0.0318 \\
    MIA, O stdev (max risk) & 0.0010 & 0.0010 & 0.0011 & 0.0204 & 0.0216 & 0.0298 & 0.0272 & 0.0126 & 0.0296 & 0.0173 & 0.0138 & 0.0024 & 0.0316 & 0.0210 & 0.0282 \\
    \midrule
    DOMIAS, O (no risk) & 0.0040 & -0.0066 & 0.0052 & 0.0966 & 0.0966 & 0.0120 & 0.0162 & -0.0088  & 0.0074 & -0.0014 & -0.0016 & -0.0054 & 0.0008 & -0.0018 & 0.0\\
    DOMIAS, O (max risk) & 0.5368 & 0.0312 & 0.3152 & 0.3542 & 0.1444 & 0.2202 & 0.2154\textsuperscript{*} & -0.0016  & 0.0136 & 0.0024\textsuperscript{*} & -0.0012\textsuperscript{*} & 0.0034\textsuperscript{*} & 0.0076\textsuperscript{*} & -0.0018\textsuperscript{*} & 0.0\\
 
    \midrule
    Link. (no risk)     & 0.0015 & 0.0116 & 0.0030 & 0.0004 & 0.0111 & 0.0035 & 0.0015 & 0.0015 & 0.0065 & 0.0 & 0.0025 & 0.0005 & 0.0015 & 0.0010 & 0.0035 \\
    Link. (max risk)    & 0.6433 & 0.9934 & 0.6336 & 0.2854 & 0.3820 & 0.2874 & 0.1589 & 0.0065 & 0.0521 & 0.0005 & 0.0045 & 0.0010 & 0.0015 & 0.0015 & 0.0020\\
    Link. stdev (max risk) & 0.0211 & 0.0035 & 0.0213 & 0.0199 & 0.0218 & 0.0203 & 0.0161 & 0.0049 & 0.0103 & 0.0017 & 0.0059 & 0.0024 & 0.0025 & 0.035 & 0.0046 \\
    \midrule 
    AIA, ML (no risk)     & 0.1570 & 0.1904 & 0.0052 & 0.0077 & 0.3370 & 0.0060 & 0.0887 & 0.6611 & 0.0 &  0.0 & 0.0 & 0.0023 & 0.0 & 0.6370 & 0.0 \\
    AIA, ML (max risk)    & 0.4499 & 0.9749 & 0.1535 & 0.3107 & 0.7193 & 0.1733 & 0.2175 & 0.4377 & 0.0337 & 0.0 & 0.0 & 0.0 & 0.0488 & 0.5702 & 0.0 \\
    \midrule 
    AIA, D (no risk)     & 0.0835 & 0.1312 & 0.2153 & 0.0665 & 0.1262 & 0.2013 & 0.1086  & 0.0887 & 0.2159 & 0.0176 & 0.0177 & 0.0229 & 0.0731 & 0.0988 & 0.1720 \\
    AIA, D (max risk)    & 0.9922 & 0.9920 & 0.9579 & 0.5945 & 0.5557 & 0.6393 & 0.3621 & 0.1161 & 0.2857 & 0.0241 & 0.0229 & 0.0244 & 0.0923 & 0.1112 & 0.1968 \\
    AIA, D stdev (max risk) & 0.0041 & 0.0080 & 0.0390 & 0.0551 & 0.0599 & 0.1221 & 0.0819 & 0.0887 & 0.1629 & 0.0585 & 0.0963 & 0.0908 & 0.1348 & 0.0466 & 0.1647 \\
    \midrule
    GTCAP (no risk)     & 0.0019 & 0.0005 & 0.0015 & 0.0089 & 0.0    & 0.0012 & 0.0817 & 0.0 & 0.0424 &  0.0 & 0.0 & 0.0 & 0.0933 & 0.0 & 0.0001 \\
    GTCAP (max risk)    & 0.9665 & 1.0    & 0.9897 & 0.5094 & 0.0156 & 0.8114 & 0.4513 & 0.0 & 0.1516 &  0.0 & 0.0 & 0.0 & 0.1303 & 0.0 & 0.0017 \\

    \bottomrule
  \end{tabular}}
  
\end{sidewaystable*}

\begin{table*}
\footnotesize
  \caption{Mean of computation times of the various measurements in seconds for the RealTabFormer and AIM models}
  \label{tab:resTime}
  \begin{tabular}{c|ccc|ccc|ccc}
    \toprule
                         & \multicolumn{3}{c|}{{\bf Leaky}}                & \multicolumn{3}{c|}{{\bf Overfit}}             & \multicolumn{3}{c}{{\bf DP}}\\
    {\bf Method}         &  {\bf Adult}  &  {\bf Texas}   &  {\bf Census} &  {\bf Adult}  &  {\bf Texas}  &  {\bf Census} &  {\bf Adult}  &  {\bf Texas}   &  {\bf Census}\\\midrule
    IMS & 22.50 & 66.30 & 43.41 & 22.50 & 66.30 & 43.41 & 22.50 & 66.30 & 43.41 \\
    DCR & 6.91 & 134.72 & 83.35 & 6.91 & 134.72 & 83.35 & 6.91 & 134.72 & 83.35 \\
    $k$-NN & 6.91 & 134.72 & 83.35 & 6.91 & 134.72 & 83.35 & 6.91 & 134.72 & 83.35 \\\midrule
    DOMIAS (O) & 2.11 & 46.45 & 68.45 & 2.21 & 45.71 & 65.45 & 2.90 & 47.86 & 71.45\\
    MIA (D) & 6.91 & 134.72 & 83.35 & 6.91 & 134.72 & 83.35 & 6.91 & 134.72 & 83.35 \\\midrule
    Link. & 18.06 & 103.01 & 99.26 & 12.44 & 30.02 & 63.50 & 9.72 & 28.74 & 63.53 \\\midrule
    AIA (ML) & 0.35 & 0.58 & 0.45 & 0.36 & 0.54 & 0.43 & 0.036 & 0.55 & 0.46\\
    AIA (D) & 163.23 & 1380.70 & 1603.49 & 74.98 & 438.33 & 1117.48& 71.99 & 445.90 & 1278.39\\
    GTCAP & 417.05 & 2093.73 & 5580.93 & 420.77 & 2115.10 & 5488.23 & 435.16 & 2238.01 & 5789.22\\

    \bottomrule
  \end{tabular}
\end{table*}




\section{Discussion}
We note that two of the generative models in our experiments failed to offer proper utility, namely: PATEGAN, and AIM, (see Appendix~\ref{app:util}). These generators failed to capture the distributions of their input data, and may have low privacy risks due to their lack of realism. We restrict our discussion to the results of the other generators.

\subsection{Quantification approaches}
Based on the taxonomy studied in Section~\ref{sec:quant}, we identified three main approaches to privacy risk quantification for synthetic data as an anonymization method: formal properties; statistical indicators; and attack simulations. Out of these only attack simulations involve explicitly modeling adversaries' threat models. This allows for privacy measurement to involve properties of both the synthetic datasets and of their generators. On the other hand, relying on specific assumptions (on model access, available auxiliary information, attack vectors, etc.) may lead to metrics that do not generalize well to other scenarios and future developments, misaligning with the Reasonableness and Proportionality Principle (cf. Section~\ref{subs:RPP}). This is particularly true for black-box and white-box attacks, which additionally tend to involve unreasonable assumptions in terms of computational resources.\\
\indent On the other hand, DP and statistical privacy indicators are ``threat model-agnostic'': their risk quantification is independent of possible risk exploitations in specific attacks. However, our results indicate that statistical privacy metrics are strongly correlated with metrics based on specific attacks in both idealized and practical risk scenarios. Furthermore, the statistical metrics showcased more robustness (see the standard deviations in Table~\ref{tab:res}), and are considerably more computationally efficient than attack-based methods (Table~\ref{tab:resTime}). Experiments with $k$-NN metrics consistently indicate that the DCR (i.e. the case $k=1$) provides the most stringent risk indication (Appendix~\ref{app:kValues}). This is consistent with prior research (see~\cite{choi2018generating, Goncalves}) and may explain why the choice for $k=1$ is common in distance-based attacks (e.g. in~\cite{TAPAS, Anonymeter}).

Under the \textbf{leaky risk model}, most metrics increase linearly with $f_l$, indicating a proportional response to the amount of direct leaks. Furthermore, with few exceptions, measures $M$ satisfy $M\approx f_l$ throughout, indicating that all of the injected risk is detected. In particular, Table~\ref{tab:res} shows that most metrics (nearly) achieve their maximum value when $f_l=1$, except for the linkability attack, the ML-based AIA and DOMIAS. A possible reason why linkability does not attain its largest value is that its efficacy depends on the partitioning of attributes. If partial records have many duplicates, linking remains difficult even with significant leaks. For the ML-based AIA, the behavior is due to the strong performance of the ML models on the control set. The baseline computation thus reduces the overall value of the metric. The efficacy of the DOMIAS attack hinges on two factors: the density estimator's sophistication and the statistical separation between member and non-member data. A simple estimator may fail to detect the local density spikes caused by model overfitting, leading to a low DOMIAS score due to poor separability. Conversely, if non-member points are highly similar to the training points, occupying the same high-density regions, the computed densities for both groups overlap significantly. This overlap also results in a low DOMIAS score, which, in this context, is interpreted as the synthetic dataset offering strong privacy protection. Thus, a high DOMIAS score requires both a sensitive estimator and discernible differences between member and non-member densities.

Under the \textbf{overfitting risk model}, responses to risk between the various methods again follow a similar pattern, resulting in large correlations. With the maximum amount of risk inserted during the experiments ($f_o=2$). Thus, even with an overfit generator whose validation loss is about twice its minimum, measured risks are about half as severe as in the more theoretical worst-case of the leaky risk model. Correlations between metrics remain strong, but weaker than under the leaky risk model, particularly on the Texas dataset. The relation between overfitting and detected risk underscores that overfitting leads to risks in synthetic data generation. This can be used pro-actively to foster privacy protection through early stopping techniques during generator training.

Under the \textbf{differential privacy risk model}, there was little response to an increase in privacy budget. This lack of measured risks makes measured correlations uninterpretable and is likely due to the aforementioned lack of utility of the DP generators' output (see Appendix~\ref{app:util}).

\paragraph{Broader implications for anonymization.}
Taken together, the experimental results underscore that the anonymization status of synthetic data is an empirical matter. It depends on measurable properties of the generator and its outputs rather than on assumptions about synthetic data as such. Consequently, privacy assessment must be dataset-specific and sensitive to both model behaviour and underlying data characteristics. The findings also show that privacy risks may persist even where utility is low, and that high utility does not necessarily imply unacceptable risks. Ensuring alignment with the GDPR’s identifiability standard therefore requires explicit verification that no metric detects leakage consistent with information “relating to” identifiable individuals in content, rather than reliance on general statistical similarity alone.

\subsection{Risk factors and attack vectors}
The taxonomy studied in Section~\ref{sec:quant} also indicates that uniqueness and similarities between real and synthetic records are core privacy risk factors.

\subsubsection{Uniqueness}
Unique records in real-world datasets are naturally more susceptible to singling out as defined by WP29 (see Section~\ref{sec:WP29ao}). Uniqueness is therefore a risk factor, regardless of the anonymization technique. Also, uniqueness is therefore an attack vector in various attack-based privacy metrics, both for singling out and AIA (see Table~\ref{tab:attr}). Statistical privacy metrics typically do not address uniqueness, while DP operates by limiting impact of individual records, thus reducing uniqueness. Interestingly, only the uniqueness-based MIA showed some response to DP privacy budget on the Texas dataset, though this was not very pronounced.\\
\indent To document the impact of uniqueness on privacy metrics, we conducted experiments with generators trained on original datasets with removed outlier fractions. We did so with both no and considerable deliberate overfitting ($f_o=1$ and $f_o=1.6$, respectively), see Appendix~\ref{app:outliers}.
For the \emph{Texas dataset}, outlier removal had a significant impact on most attacks, which struggled to succeed after the outliers were removed, especially under considerable overfitting ($f_o=1.6$). On the \emph{Adult} and \emph{Census} datasets, only the DOMIAS attack showed a significant degradation in attack efficacy after outlier removal. The attacks likely succeeded by exploiting the model's highly specific generations for unique, easily identifiable outlier data points. Consequently, after these outliers were removed from the training data, the attacks failed, as their most obvious targets were gone.\\
\indent These results corroborate the theory that outliers pose significant privacy hazards, potentially arising from local overfitting around these unique points~\cite{carlini2019secret, frolich2024fraunhofer}. The results indicate that this particularly affects MIAs, which is intuitive as these typically identify unique individuals to target. This finding is also consistent with the MIA literature (see, e.g.~\cite{groundhog, meeus2023achilles, yao2025dcr, vanbreugel2023DOMIAS}). The DCR, by contrast, was able to detect risks robustly even after outlier removal. This is evidenced by the consistently higher DCR value with overfit model than with non-overfit model, regardless of the proportion of outliers removed.\\
\indent Our results indicate that dataset properties can affect the extent to which outlier removal mitigates privacy risks. The \emph{Texas} dataset is heavily categorical (21 out of 28 attributes), with many attributes having few values. This limited combination space may make outliers more obvious targets for a wider range of attacks. Conversely, the Adult and Census sets have smaller proportions of categorical attributes (9 out of 15 and 11 out of 37, respectively) and broader value ranges, perhaps making their outliers exploitable only by highly specific methods like DOMIAS.

\subsubsection{Similarity, ML, and combined factors}
Similarity between synthetic and real records can increase susceptibility to all three WP29 attacks. For singling out, if some unique real record does not have a corresponding highly similar synthetic record, then the uniqueness does not present as an attack vector. This is corroborated by our experimental results: uniqueness-based attacks and distance-based metrics are strongly correlated (Appendix~\ref{app:full}); and outlier removal reduces risks measured through similarity-based metrics for the Texas dataset (Appendix~\ref{app:outliers}). Thus, similarity-based metrics can provide direct insight into uniqueness as a risk factor.\\
\indent Similarity facilitates linkability, as it enables linking synthetic records to given real records, providing insight into the linkability of the training data. For AIAs, similarity can serve as a direct attack vector even in the absence of uniqueness. However, in AIAs against non-outliers, the base-rate problem may be more pronounced. Our experiments with canary records indicate that this problem is significant for non-outliers and may result in an overestimation of AIA risks (Appendix~\ref{app:canary}), particularly when adhering to the legal perspective that only ``relating to in content'' poses a privacy breach. This was particularly true for the GTCAP, for which the canary baseline experiment indicated that nearly all of the success could be ascribed to the base-rate problem. \\ 
\indent Our results also indicate that distance-based privacy metrics strongly correlate with no-box attacks that leverage ML as an attack vector. This has an intuitive explanation, as classifiers or regressors will assign similar values to similar records. In the black-box case, Meeus et al.~\cite{meeus2023achilles} show how exploiting the combination of uniqueness and similarity can make MIAs possible that would otherwise not meet the reasonableness principle (Section~\ref{subs:RPP}) due to their computational requirements.

\subsubsection{Implications for GDPR-Aligned Risk Assessment}
The combined results across uniqueness-, similarity-, and ML-based vectors show that privacy risks in synthetic data depend on structural features of the dataset and on generator learning dynamics. Accordingly, whether synthetic data qualify as anonymized under the GDPR hinges on empirically demonstrating the absence of detectable leakage under threat models aligned with the “means reasonably likely to be used” standard. This reinforces a risk-based, rather than assumption-based, understanding of synthetic data as a privacy-enhancing technology.

\section{Conclusion}
This paper proposed a framework for assessing the efficacy of tabular synthetic data privacy metrics. To showcase the framework, it first provided a survey of existing privacy metrics, covering mathematical privacy properties; statistical privacy indicators; and simulated attacks. The further classification of the latter into singling out, linkability, and inference attacks links them to relevant legal theory on anonymization. 

Our framework evaluates metrics from the taxonomy across three novel \emph{risk models}. This enables the empirical measurement of metrics' responses to deliberately added risk. The risk models respectively insert risk through: \begin{enumerate*}
    \item direct leaks;
    \item generator overfit;
    \item varying privacy budgets for DP generators.
\end{enumerate*} 
Through baseline computations, it covers the discrepancy between two types of information inferred by generative models: specific information about the generator's training data; and general information about the underlying population.

Experiments with the framework showed that no-box privacy quantification methods correlate strongly, indicating that uniqueness-based and similarity-based risks coincide under no-box risk models. This suggests that the choice of a privacy quantification method might be best informed by robustness and efficiency, which would favor statistical indicators. However, such indicators only explicitly target distances between real and synthetic data points. Future research should study the possibility to reliably, robustly, and efficiently quantify uniqueness as a privacy risk factor for synthetic data at the level of probability distributions, following~\cite{Acquisto2024}.

Overall, the results indicate that synthetic data can only be regarded as anonymized when empirical evidence confirms that the generator does not reproduce information relating to identifiable individuals in content. The proposed framework contributes to this determination by translating GDPR-relevant concepts—such as singling out, linkability, inference, and the “means reasonably likely to be used” standard—into measurable, dataset-specific tests. This facilitates a clearer alignment between technical evaluation practices and legal anonymization requirements, and underscores the need for multi-metric, context-aware assessment strategies rather than reliance on any single indicator or on assumptions about synthetic data’s inherent privacy properties.

\bibliographystyle{unsrt}  
\bibliography{References}

\appendix

\section{Complete experimental results}\label{app:full}


\begin{figure}[H]  
\subfloat
  {\includegraphics[width=.3\linewidth]{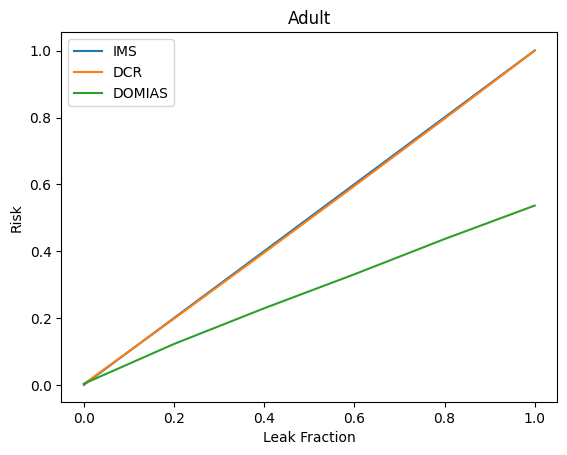}}\hfill
\subfloat
  {\includegraphics[width=.3\linewidth]{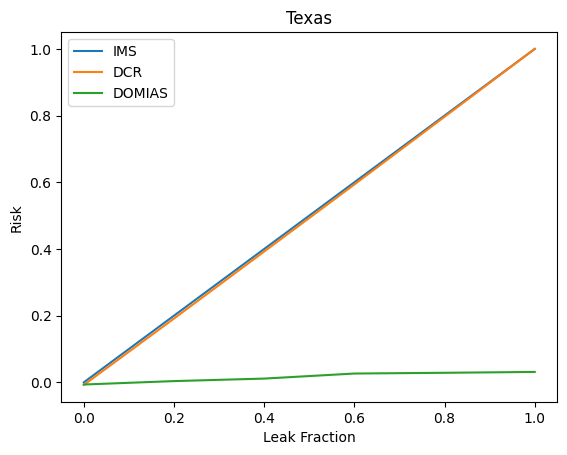}}\hfill
\subfloat
  {\includegraphics[width=.3\linewidth]{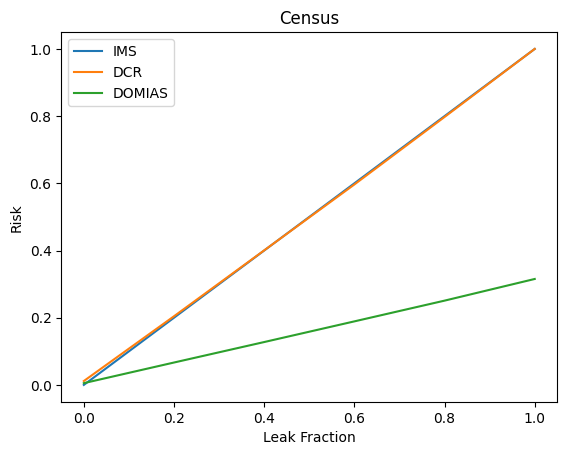}} \\

\subfloat
  {\includegraphics[width=.3\linewidth]{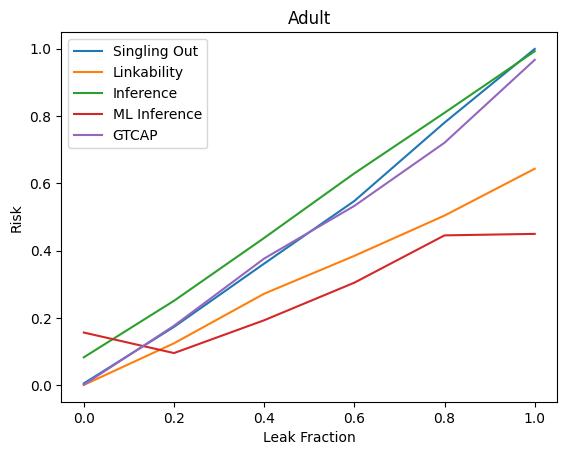}}\hfill
\subfloat
  {\includegraphics[width=.3\linewidth]{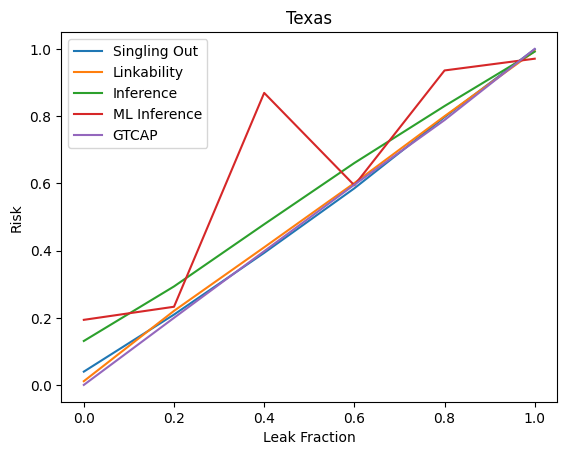}}\hfill
\subfloat
  {\includegraphics[width=.3\linewidth]{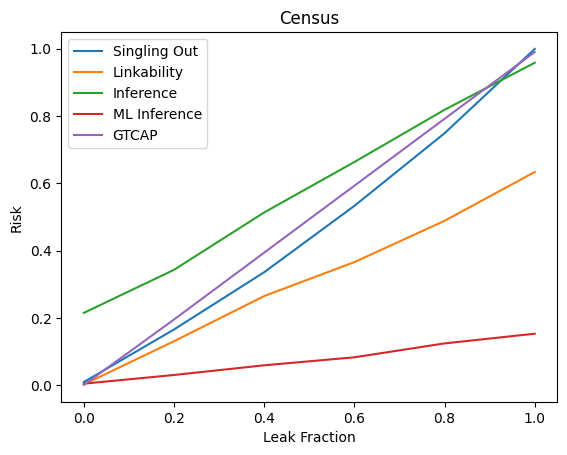}}
\caption{Risk assessment methods evaluated using the leaky risk model}\label{fig:Leak2-2}
\end{figure}


\begin{figure}[H]
\subfloat
  {\includegraphics[width=.3\linewidth]{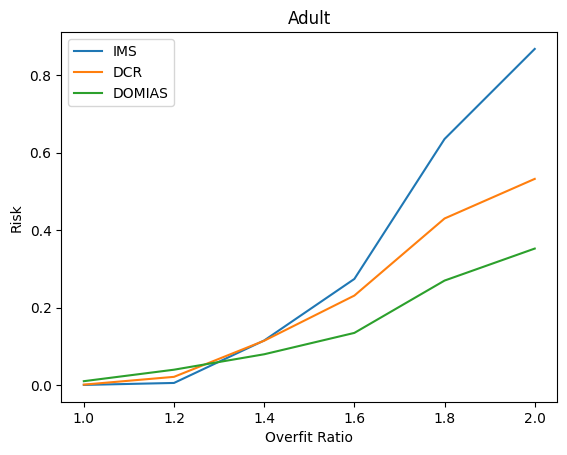}}\hfill
\subfloat
  {\includegraphics[width=.3\linewidth]{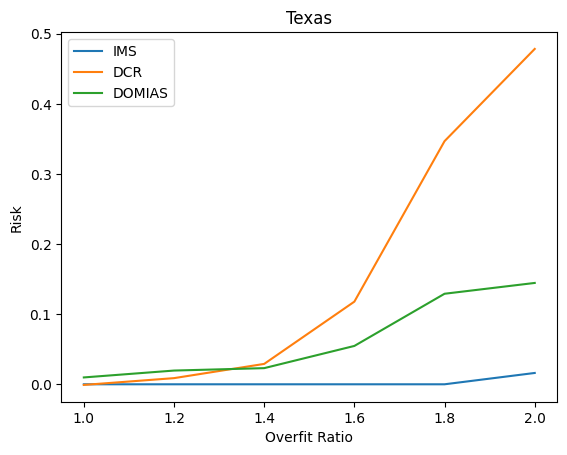}}\hfill
\subfloat
  {\includegraphics[width=.3\linewidth]{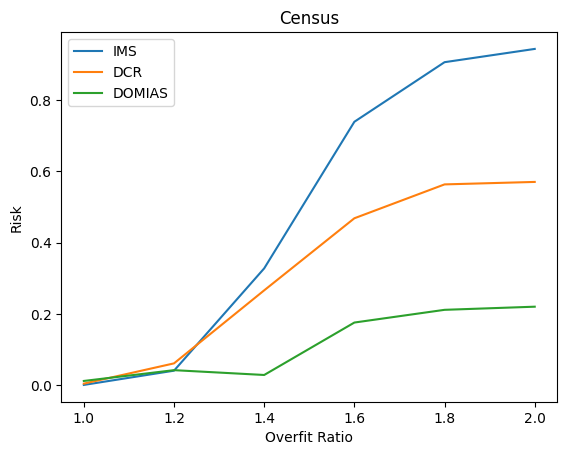}}

\subfloat
  {\includegraphics[width=.3\linewidth]{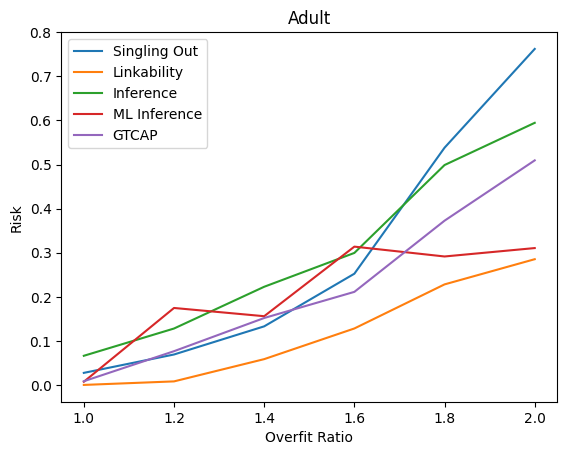}}\hfill
\subfloat
  {\includegraphics[width=.3\linewidth]{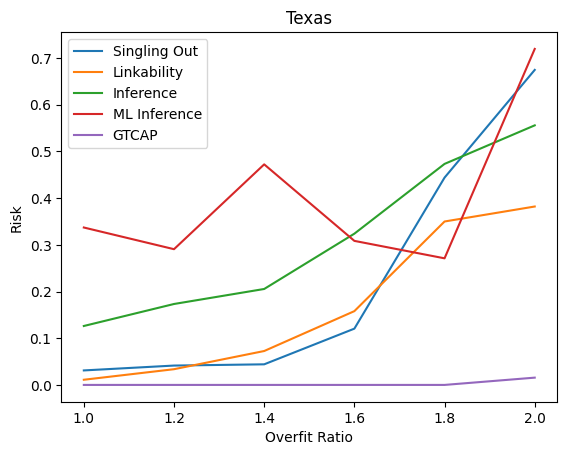}}\hfill
\subfloat
  {\includegraphics[width=.3\linewidth]{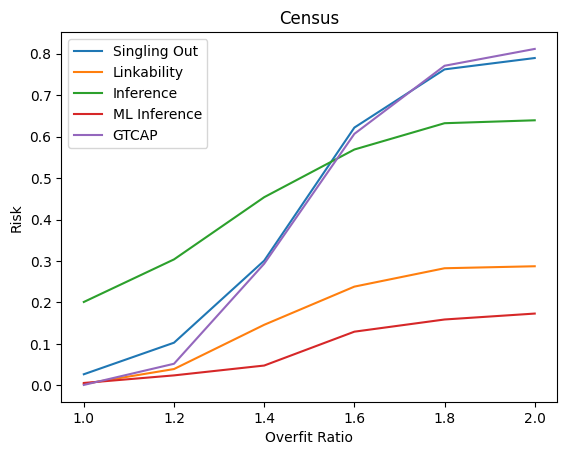}}
\caption{Risk assessment methods evaluated using the overfit risk model - RTF~\cite{solatorio}}\label{fig:over-2}
\end{figure}

\begin{figure}[H]
\subfloat
  {\includegraphics[width=.3\linewidth]{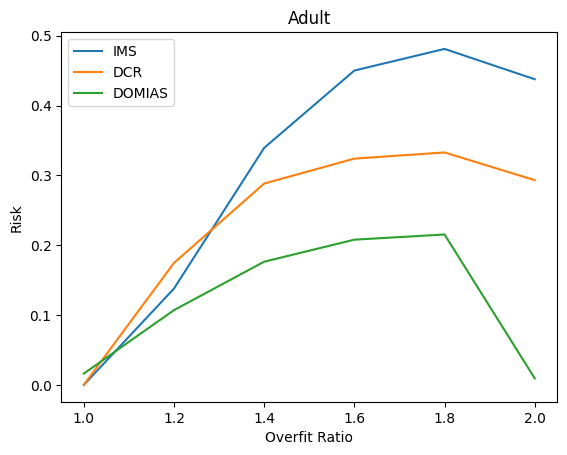}}\hfill
\subfloat
  {\includegraphics[width=.3\linewidth]{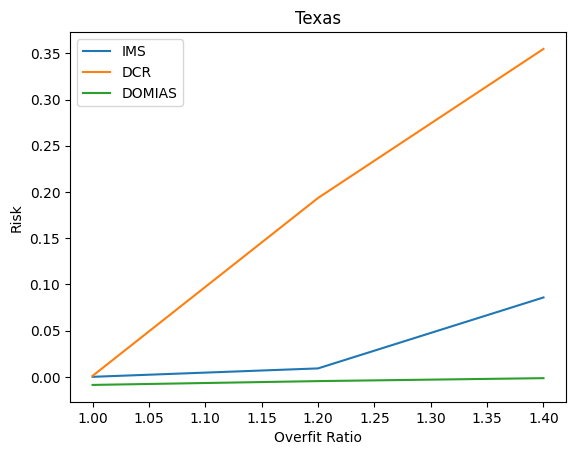}}\hfill
\subfloat
  {\includegraphics[width=.3\linewidth]{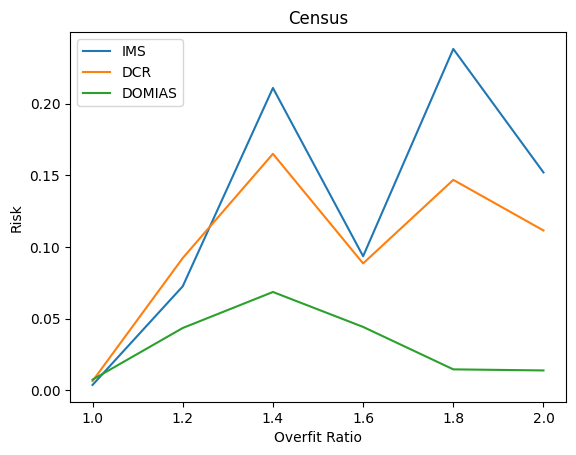}}

\subfloat
  {\includegraphics[width=.3\linewidth]{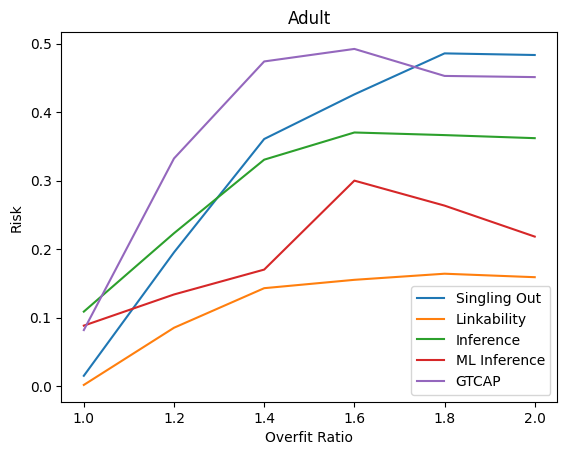}}\hfill
\subfloat
  {\includegraphics[width=.3\linewidth]{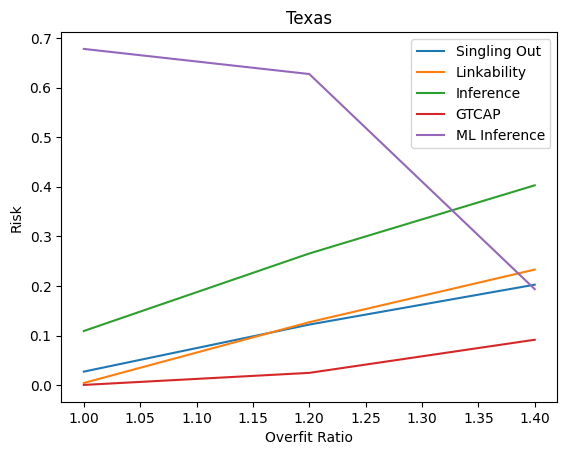}}\hfill
\subfloat
  {\includegraphics[width=.3\linewidth]{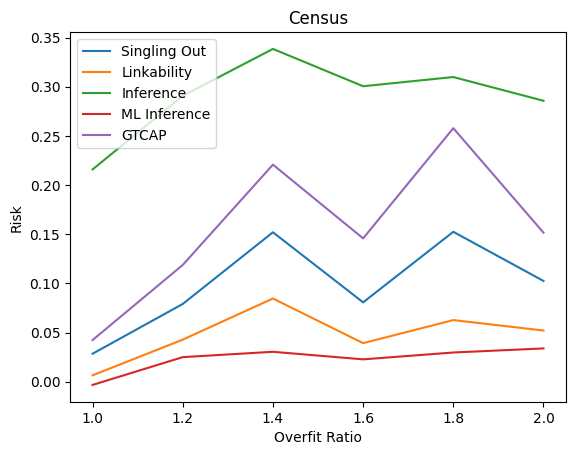}}
\caption{Risk assessment methods evaluated using the overfit risk model - Synthpop~\cite{synthpop}}\label{fig:over-2}
\end{figure}




\begin{figure}[H]
\subfloat
  {\includegraphics[width=.3\linewidth]{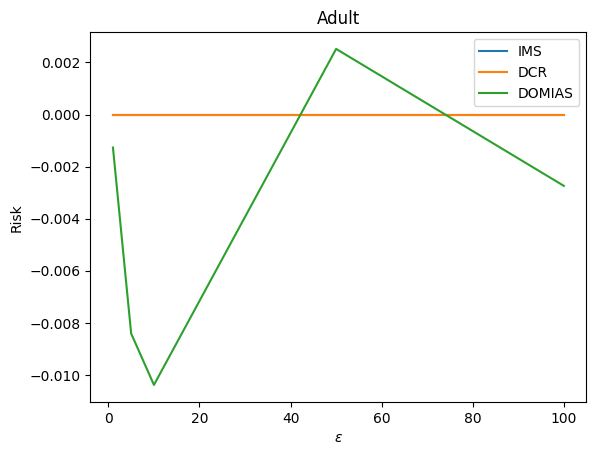}}\hfill
\subfloat
  {\includegraphics[width=.3\linewidth]{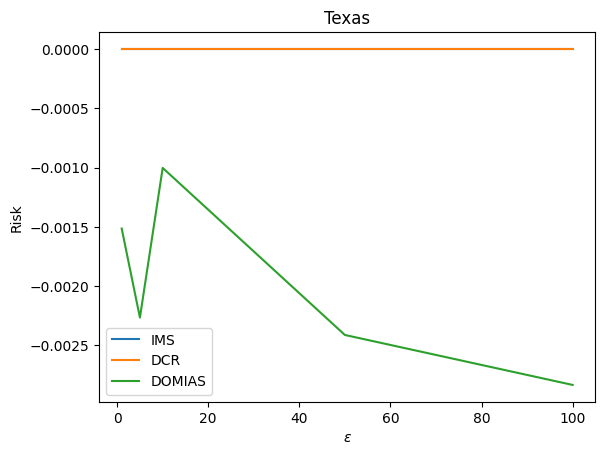}}\hfill
\subfloat
  {\includegraphics[width=.3\linewidth]{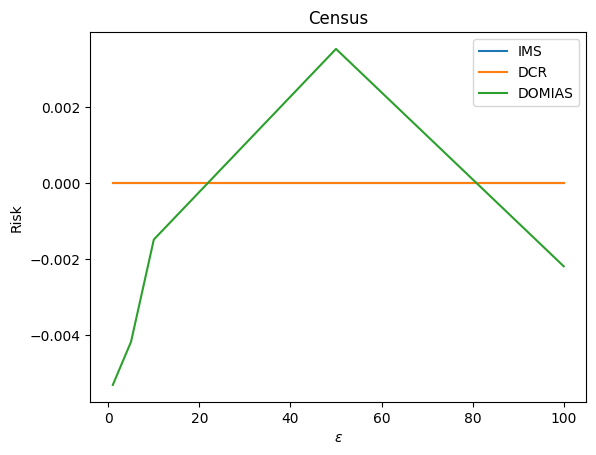}} \\

\subfloat
  {\includegraphics[width=.3\linewidth]{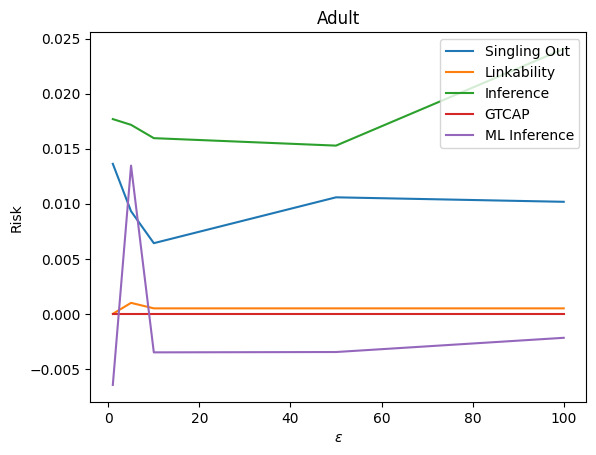}}\hfill
\subfloat
  {\includegraphics[width=.3\linewidth]{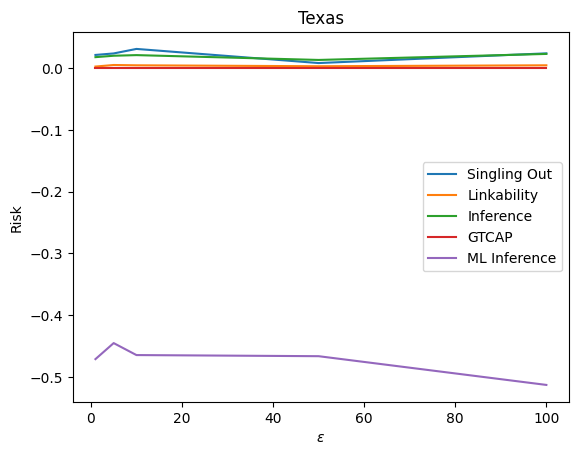}}\hfill
\subfloat
  {\includegraphics[width=.3\linewidth]{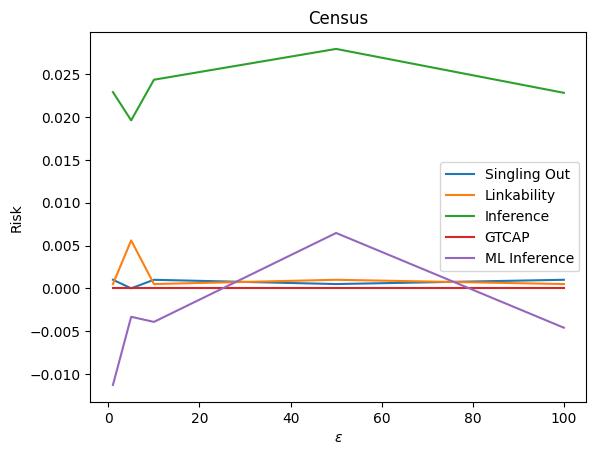}}
\caption{Risk assessment methods evaluated using the DP risk model - PATEGAN~\cite{yoon2018pategan}}\label{fig:dp-2}
\end{figure}

\begin{figure}[H]
\subfloat
  {\includegraphics[width=.3\linewidth]{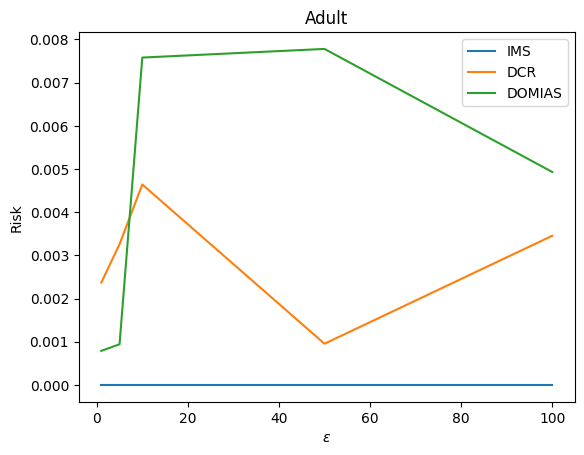}}\hfill
\subfloat
  {\includegraphics[width=.3\linewidth]{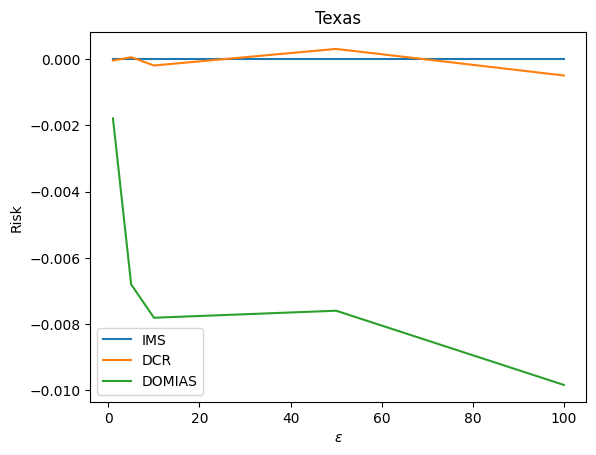}}\hfill
\subfloat
  {\includegraphics[width=.3\linewidth]{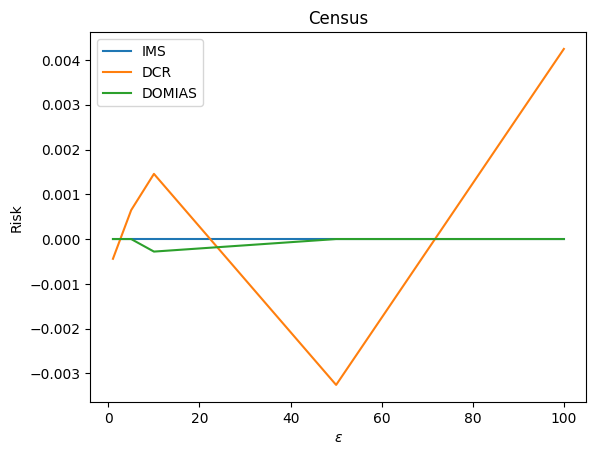}} \\

\subfloat
  {\includegraphics[width=.3\linewidth]{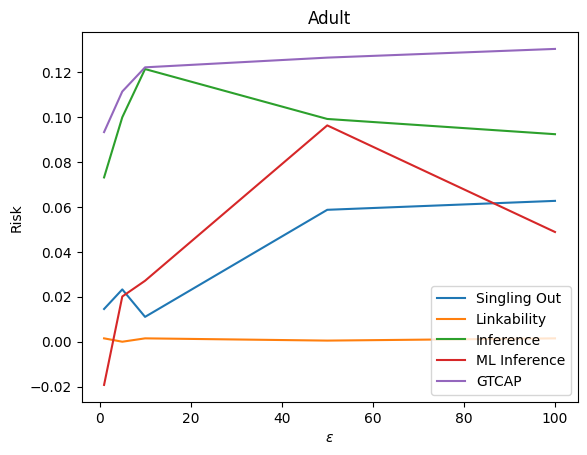}}\hfill
\subfloat
  {\includegraphics[width=.3\linewidth]{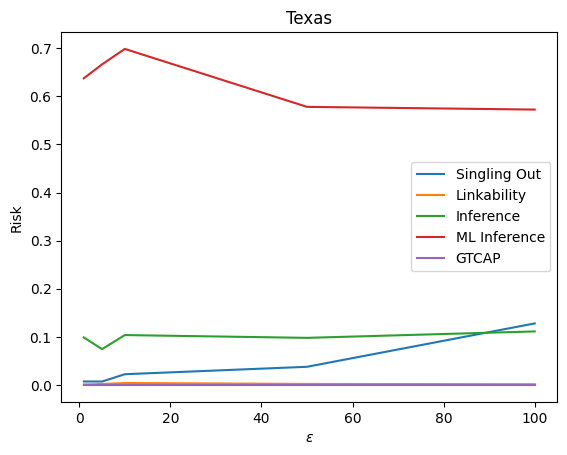}}\hfill
\subfloat
  {\includegraphics[width=.3\linewidth]{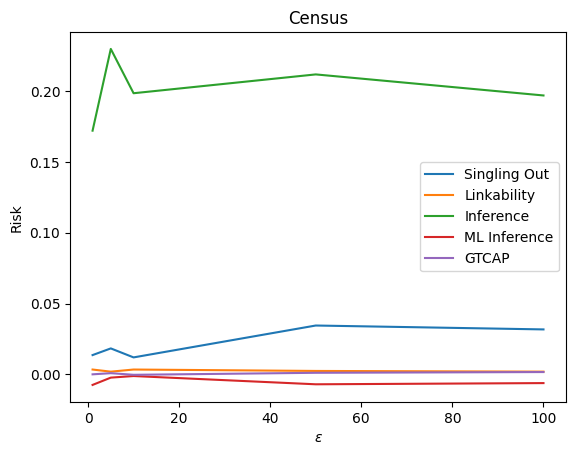}}
\caption{Risk assessment methods evaluated using the DP risk model - AIM~\cite{AIM}}\label{fig:dp-2}
\end{figure}


\begin{figure}[H]
\subfloat
  {\includegraphics[width=.3\linewidth]{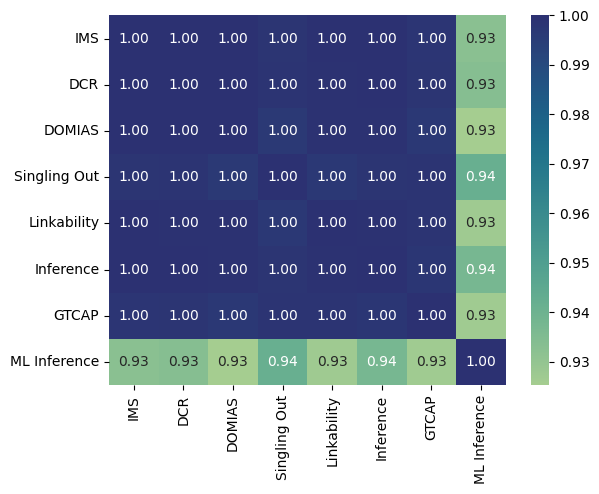}}\hfill
\subfloat
  {\includegraphics[width=.3\linewidth]{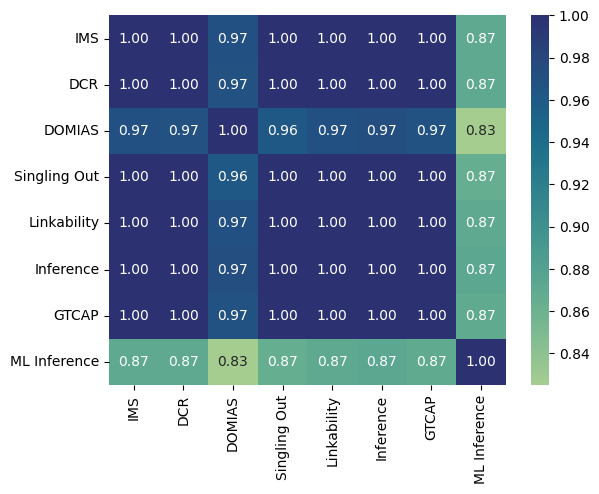}}\hfill
\subfloat
  {\includegraphics[width=.3\linewidth]{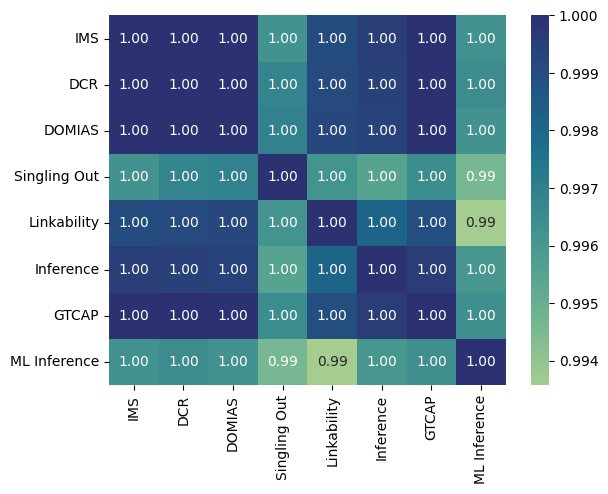}}
\caption{correlation matrices risk assessment methods using leaky risk model (from left to right: Adult, Texas, Census dataset)}\label{fig:LeakCorr2}
\end{figure}

\begin{figure}[H]
\subfloat
  {\includegraphics[width=.3\linewidth]{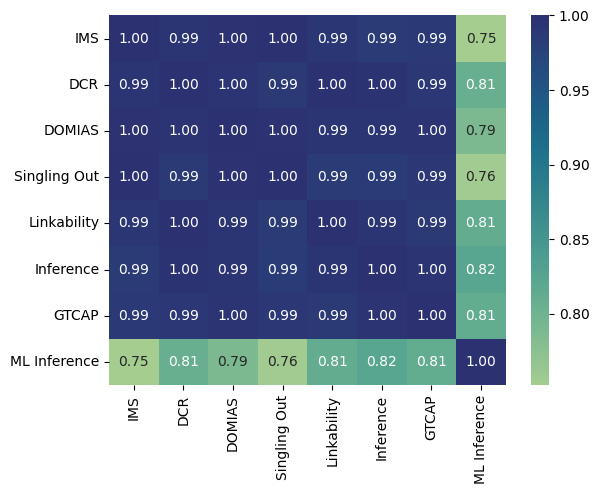}}\hfill
\subfloat
  {\includegraphics[width=.3\linewidth]{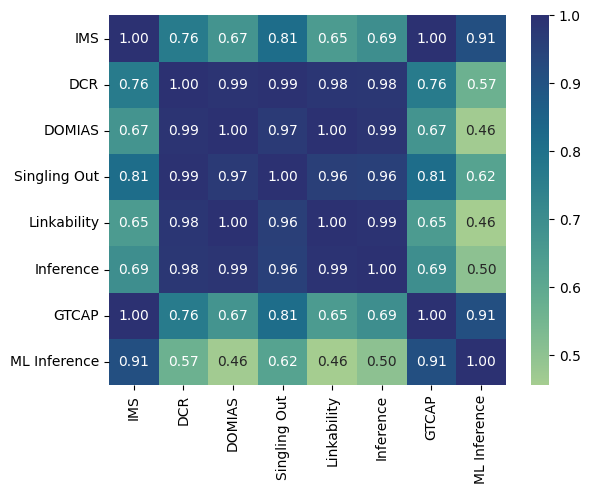}}\hfill
\subfloat
  {\includegraphics[width=.3\linewidth]{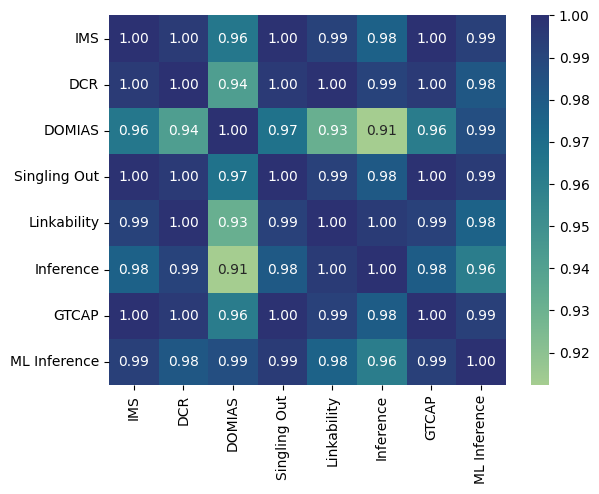}}
\caption{correlation matrices risk assessment methods using overfitting risk model - RTF(from left to right: Adult, Texas, Census dataset)}\label{fig:overfitCorr}
\end{figure}

\begin{figure}[H]
\subfloat
  {\includegraphics[width=.3\linewidth]{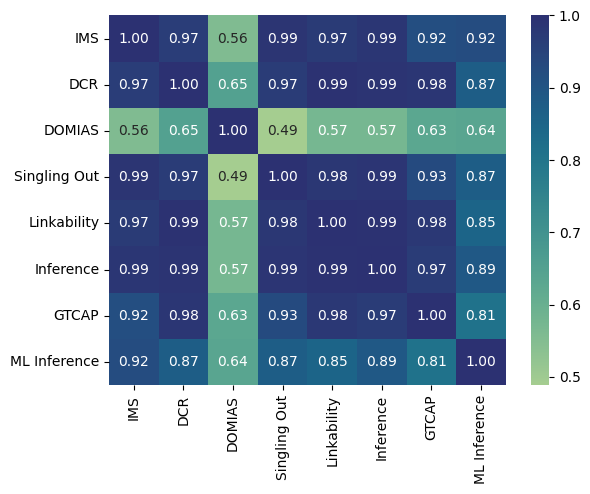}}\hfill
\subfloat
  {\includegraphics[width=.3\linewidth]{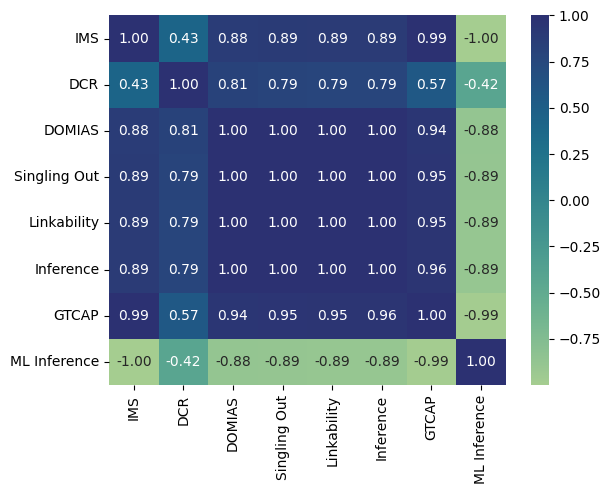}}\hfill
\subfloat
  {\includegraphics[width=.3\linewidth]{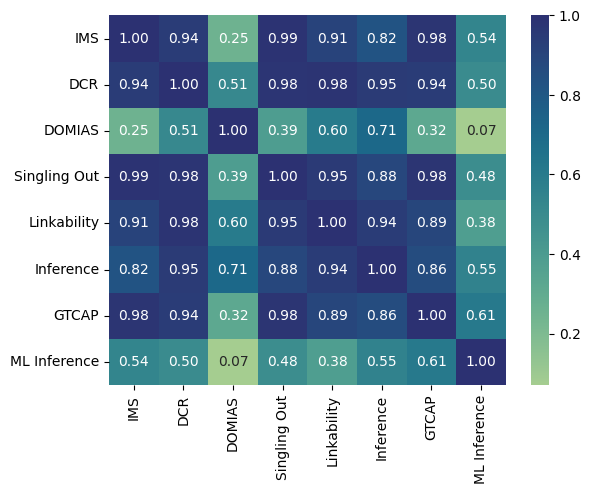}}
\caption{correlation matrices risk assessment methods using overfitting risk model - Synthpop(from left to right: Adult, Texas, Census dataset)}\label{fig:overfitCorr}
\end{figure}

\begin{figure}[H]
\subfloat
  {\includegraphics[width=.3\linewidth]{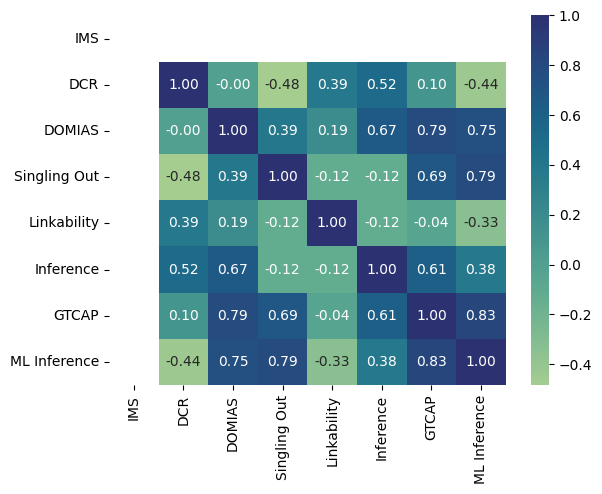}}\hfill
\subfloat
  {\includegraphics[width=.3\linewidth]{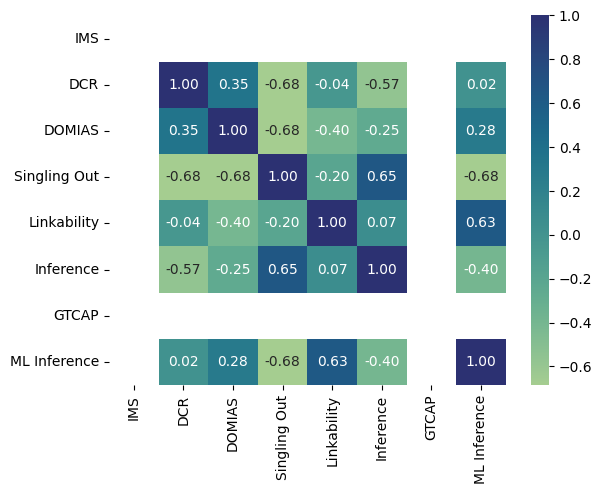}}\hfill
\subfloat
  {\includegraphics[width=.3\linewidth]{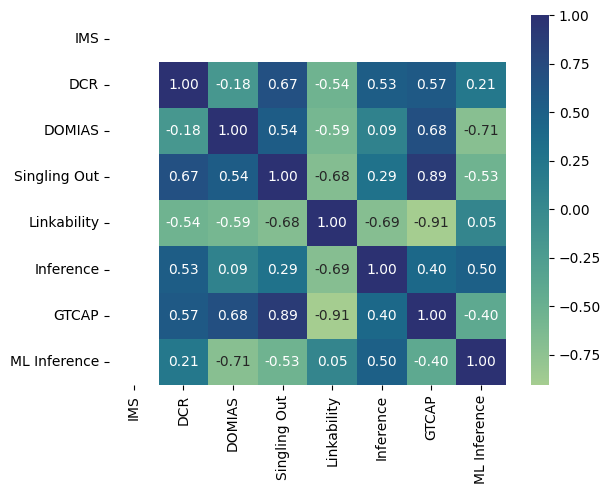}}
\caption{correlation matrices risk assessment methods using DP risk model - AIM (from left to right: Adult, Texas, Census dataset)}\label{fig:DPCorr}
\end{figure}

\begin{figure}[H]
\subfloat
  {\includegraphics[width=.3\linewidth]{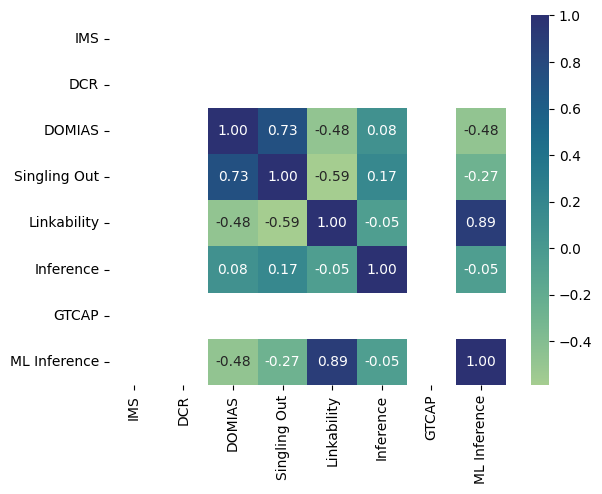}}\hfill
\subfloat
  {\includegraphics[width=.3\linewidth]{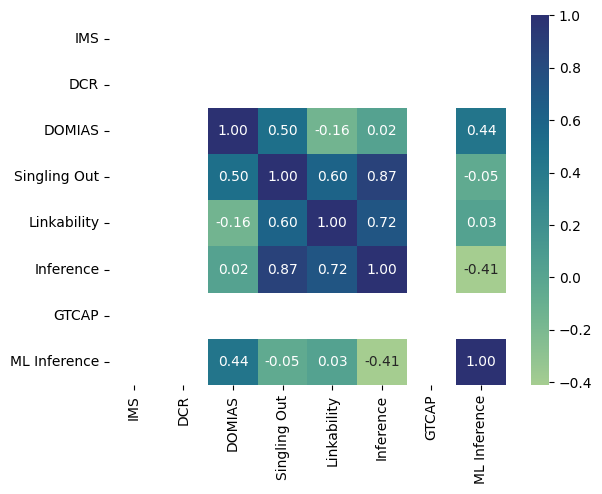}}\hfill
\subfloat
  {\includegraphics[width=.3\linewidth]{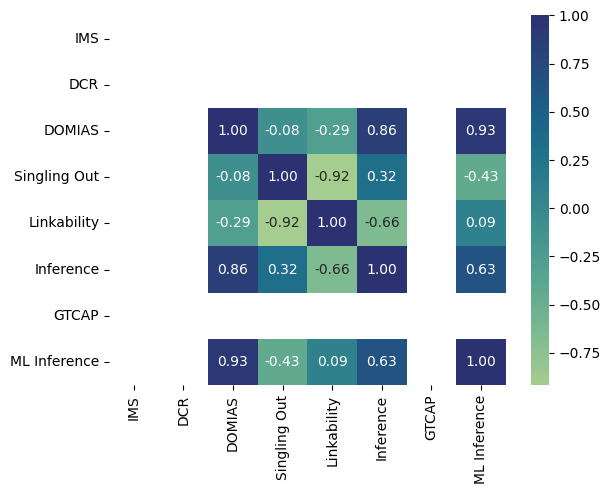}}
\caption{correlation matrices risk assessment methods using DP risk model - PATEGAN (from left to right: Adult, Texas, Census dataset)}\label{fig:DPCorr}
\end{figure}

\newpage
\section{Experiments with $k$-NN-based indicators}\label{app:kValues}
For each dataset and risk model, we experimented with $k$-NN-based privacy indicators for various values of $k$. Both the leaky and overfitting risk models show a clear dominance of $k=1$ over other values.

\begin{figure*}[h]
\subfloat
  {\includegraphics[width=.3\linewidth]{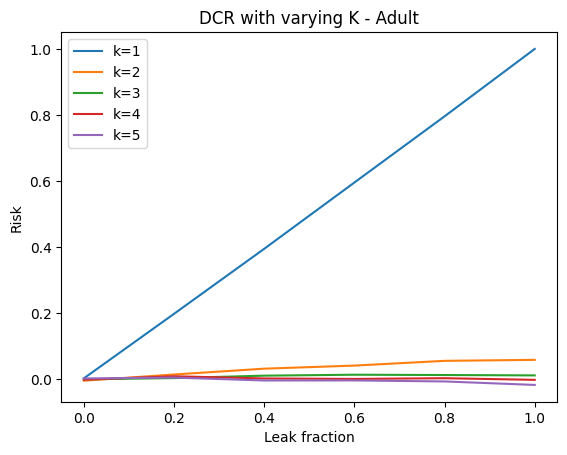}}\hfill
\subfloat
  {\includegraphics[width=.3\linewidth]{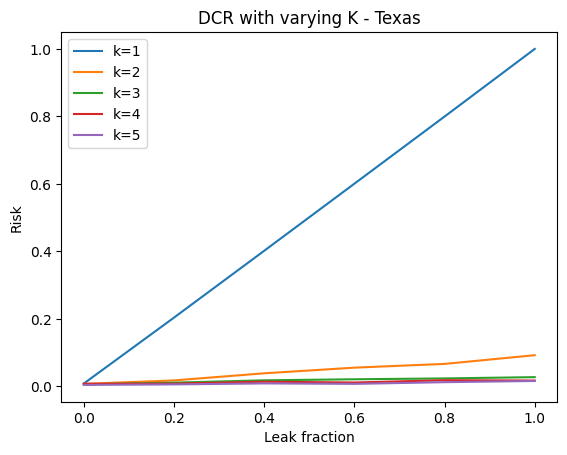}}\hfill
\subfloat
  {\includegraphics[width=.3\linewidth]{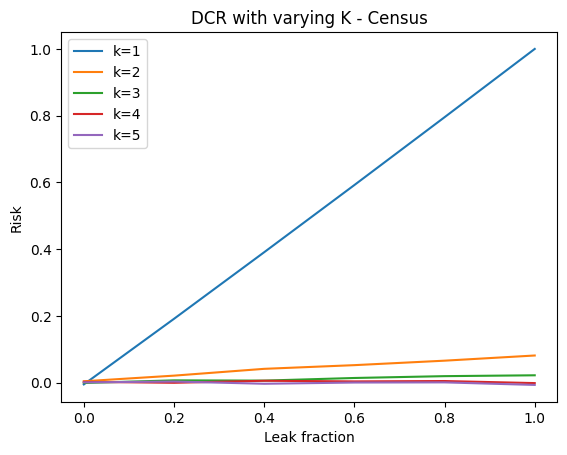}}
\caption{Results with $k$-NN-based privacy metric for various $k$ using the leaky risk model}\label{fig:LeakyKNN}
\end{figure*}


\begin{figure*}[h]
\subfloat
  {\includegraphics[width=.3\linewidth]{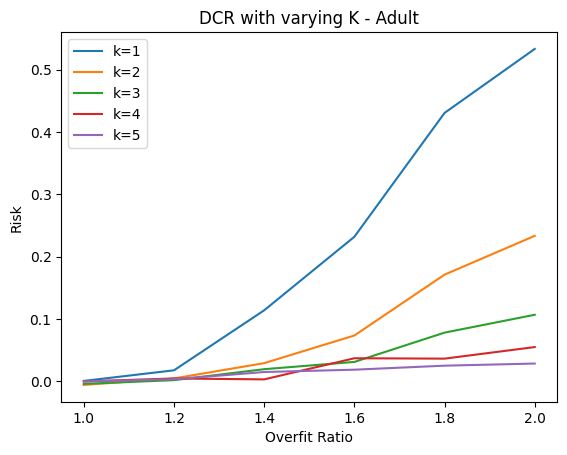}}\hfill
\subfloat
  {\includegraphics[width=.3\linewidth]{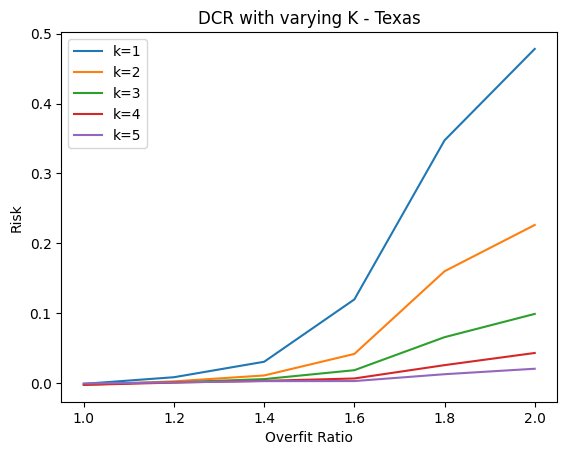}}\hfill
\subfloat
  {\includegraphics[width=.3\linewidth]{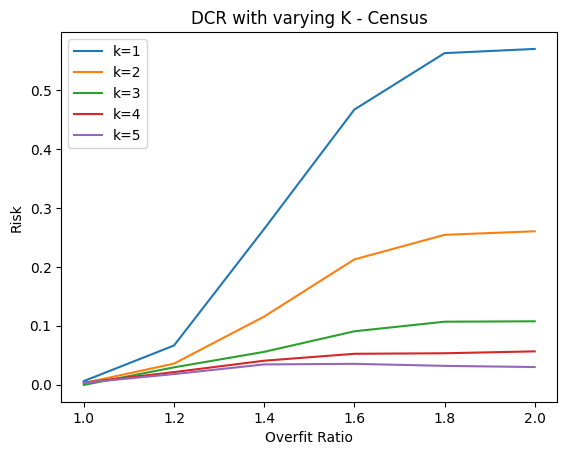}}
\caption{Results with $k$-NN-based privacy metric for various $k$ using the overfit risk model - RTF~\cite{solatorio}}\label{fig:overfitKNN}
\end{figure*}

\begin{figure*}[h]
\subfloat
  {\includegraphics[width=.3\linewidth]{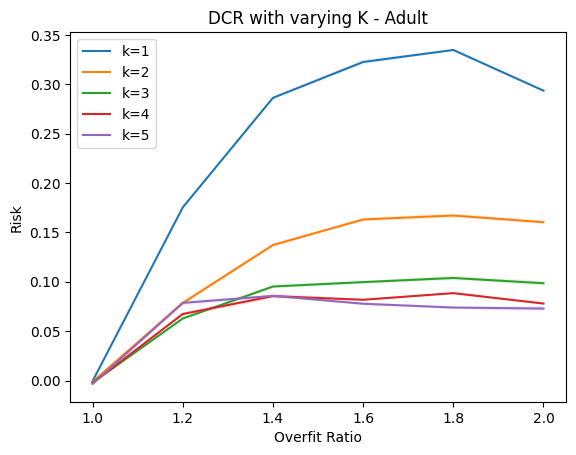}}\hfill
\subfloat
  {\includegraphics[width=.3\linewidth]{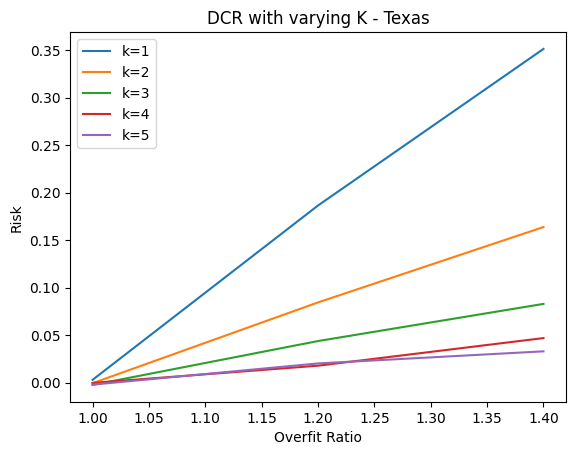}}\hfill
\subfloat
  {\includegraphics[width=.3\linewidth]{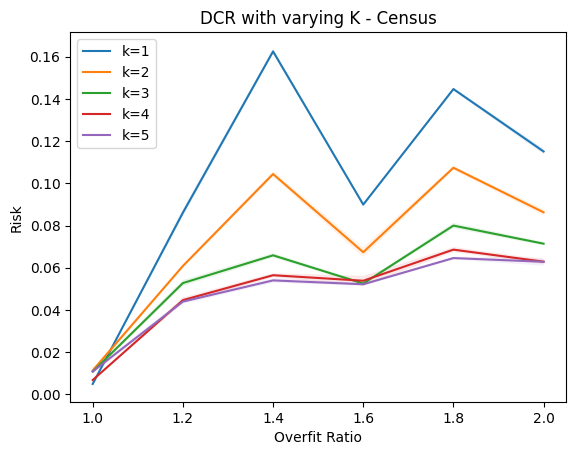}}
\caption{Results with $k$-NN-based privacy metric for various $k$ using the overfit risk model - Synthpop~\cite{synthpop}}\label{fig:overfitKNN}
\end{figure*}



\begin{figure*}[h]
\subfloat
  {\includegraphics[width=.3\linewidth]{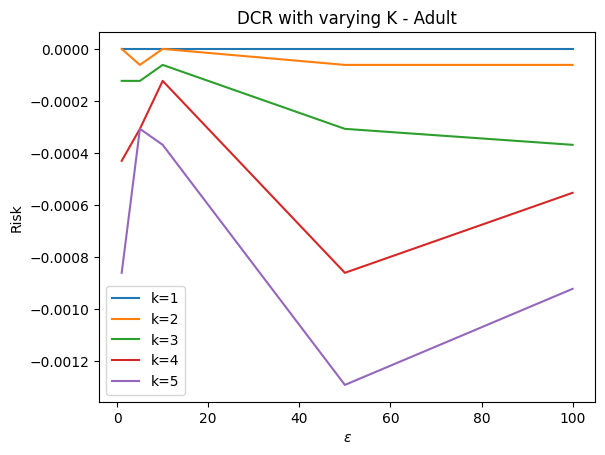}}\hfill
\subfloat
  {\includegraphics[width=.3\linewidth]{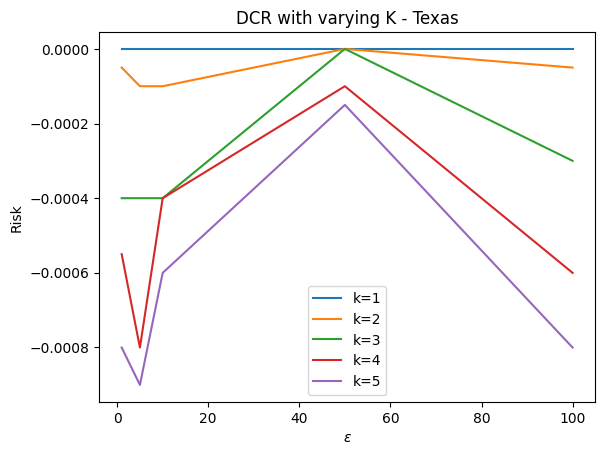}}\hfill
\subfloat
  {\includegraphics[width=.3\linewidth]{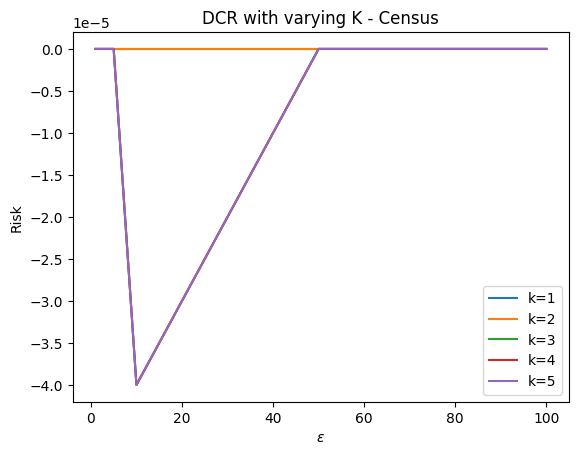}}
\caption{Results with $k$-NN-based privacy metric for various $k$ using the DP risk model - PATEGAN~\cite{yoon2018pategan}}\label{fig:overfitKNN}
\end{figure*}

\begin{figure*}[h]
\subfloat
  {\includegraphics[width=.3\linewidth]{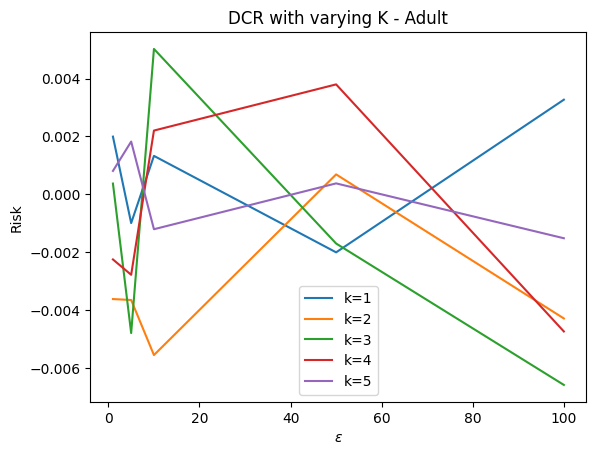}}\hfill
\subfloat
  {\includegraphics[width=.3\linewidth]{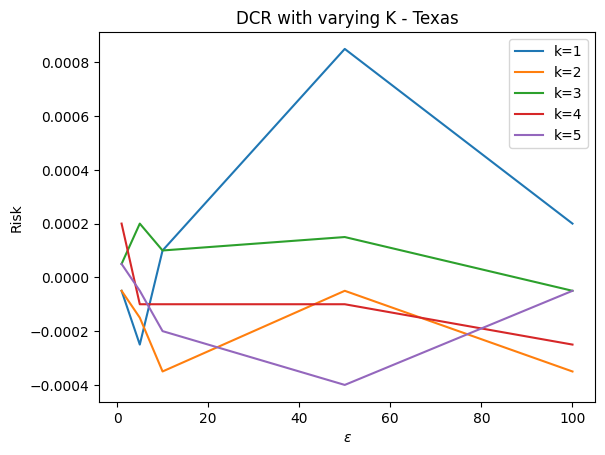}}\hfill
\subfloat
  {\includegraphics[width=.3\linewidth]{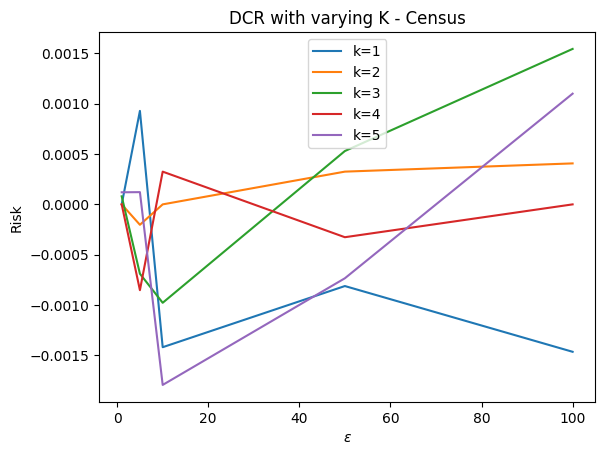}}
\caption{Results with $k$-NN-based privacy metric for various $k$ using the DP risk model - AIM}\label{fig:DPKNN}
\end{figure*}


\section{Experiments with outlier removal}\label{app:outliers}
We used the local outlier factor (LOF) to embeddings obtained through contrastive learning to identify outliers ~\cite{palacios2025contrastive}. Different proportions of outliers were removed to evaluate their impact on the measurements. We used the overfitting risk model, with overfit ratios of $1.0$ and $1.6$. 

\begin{figure*}[h]
\subfloat
  {\includegraphics[width=.3\linewidth]{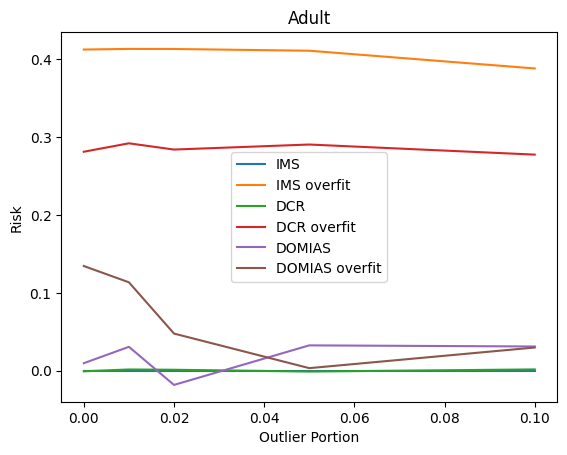}}\hfill
\subfloat
  {\includegraphics[width=.3\linewidth]{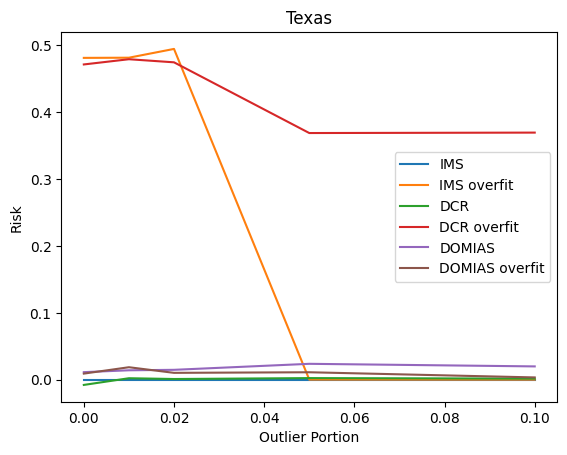}}\hfill
\subfloat
  {\includegraphics[width=.3\linewidth]{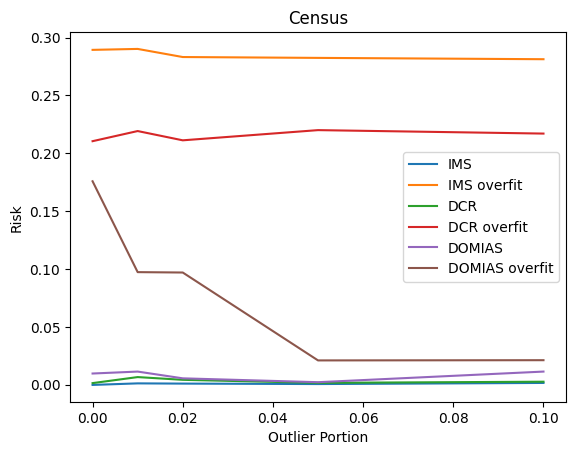}}
\caption{IMS, DCR, and MIA  with outlier removal in the original dataset prior to generator training, with both no overfitting ($f_o=1$) and overfitting ($f_o=1.6)$ - RTF~\cite{solatorio}}\label{fig:outliers-1}
\end{figure*}

\begin{figure*}[h]
\subfloat
  {\includegraphics[width=.3\linewidth]{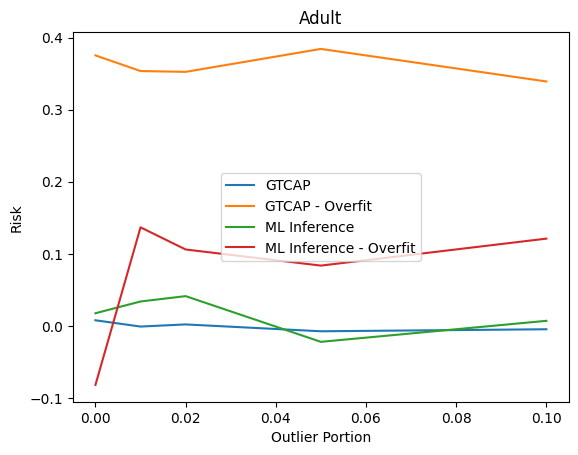}}\hfill
\subfloat
  {\includegraphics[width=.3\linewidth]{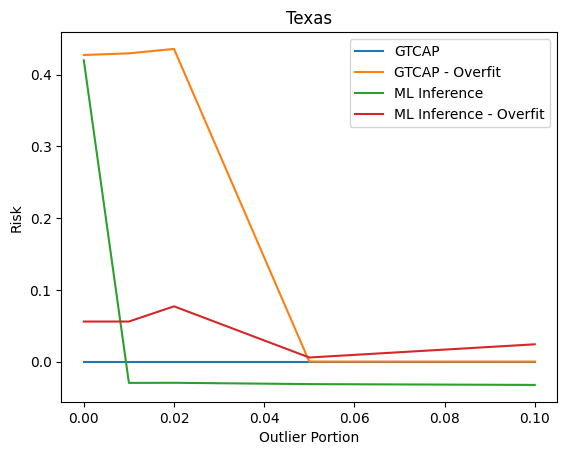}}\hfill
\subfloat
  {\includegraphics[width=.3\linewidth]{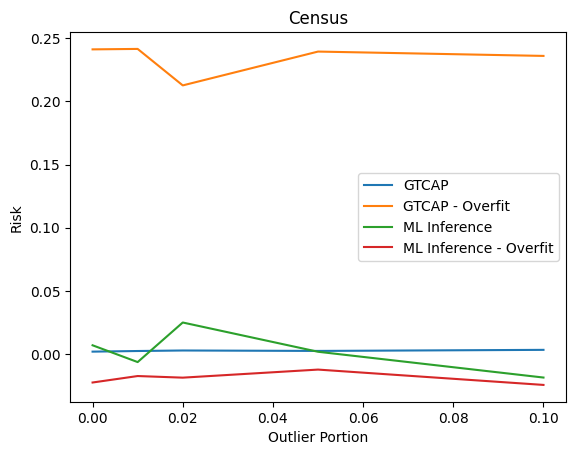}}
\caption{GTCAP and ML Inference with outlier removal in the original dataset prior to generator training, with both no overfitting ($f_o=1$) and overfitting ($f_o=1.6)$ - RTF~\cite{solatorio}}\label{fig:outliers-2}
\end{figure*}

\begin{figure*}[h]
\subfloat
  {\includegraphics[width=.3\linewidth]{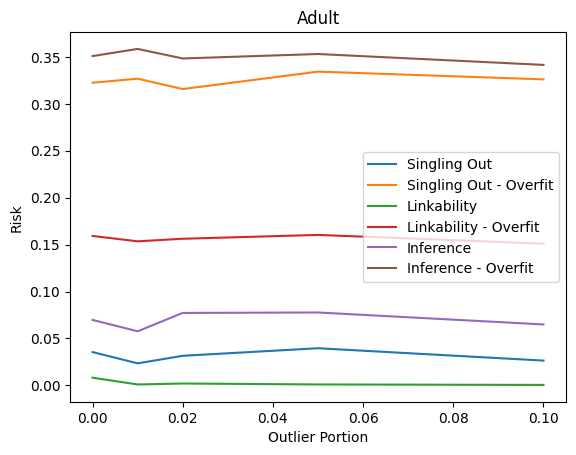}}\hfill
\subfloat
  {\includegraphics[width=.3\linewidth]{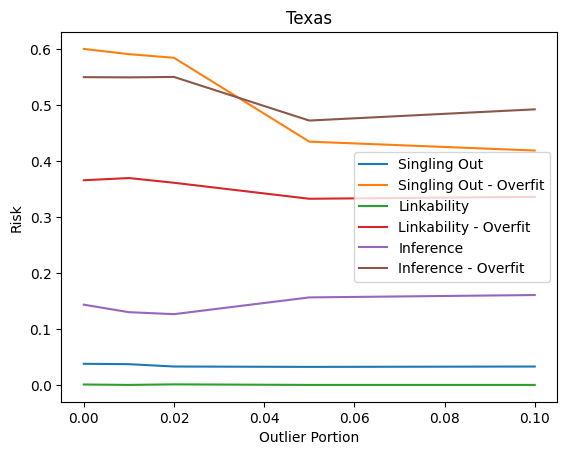}}\hfill
\subfloat
  {\includegraphics[width=.3\linewidth]{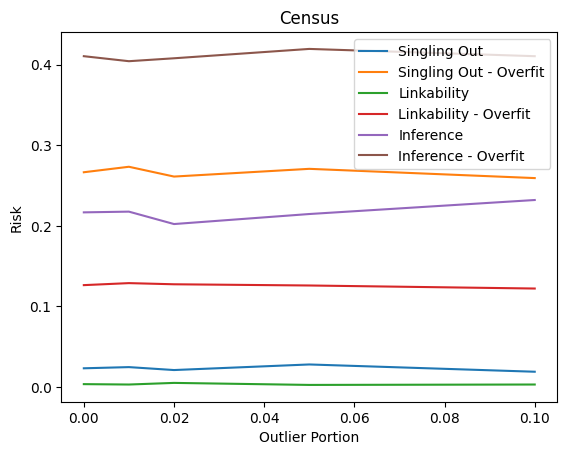}}
\caption{Anonymeter's methods with outlier removal in the original dataset prior to generator training, with both no overfitting ($f_o=1$) and overfitting ($f_o=1.6)$ - RTF~\cite{solatorio}}\label{fig:outliers-3}
\end{figure*}

\begin{figure*}[h]
\subfloat
  {\includegraphics[width=.3\linewidth]{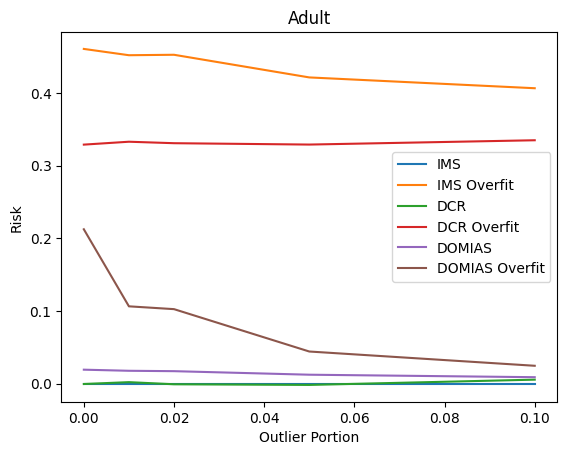}}\hfill
\subfloat
  {\includegraphics[width=.3\linewidth]{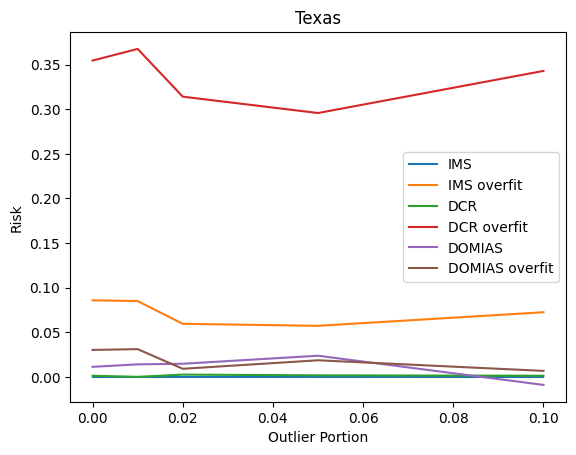}}\hfill
\subfloat
  {\includegraphics[width=.3\linewidth]{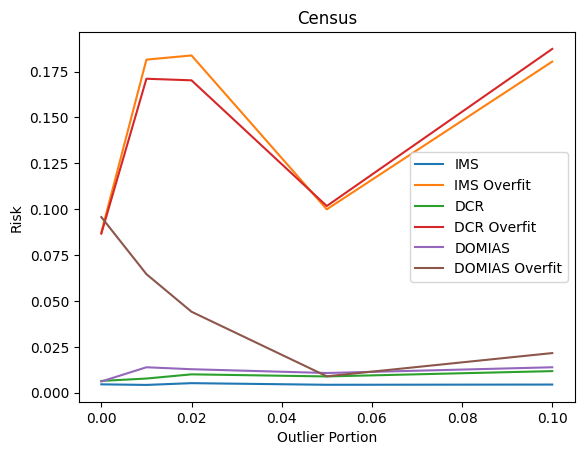}}
\caption{IMS, DCR, and MIA  with outlier removal in the original dataset prior to generator training, with both no overfitting ($f_o=1$) and overfitting ($f_o=1.6)$ - Synthpop~\cite{synthpop}}\label{fig:outliers-1}
\end{figure*}

\begin{figure*}[h]
\subfloat
  {\includegraphics[width=.3\linewidth]{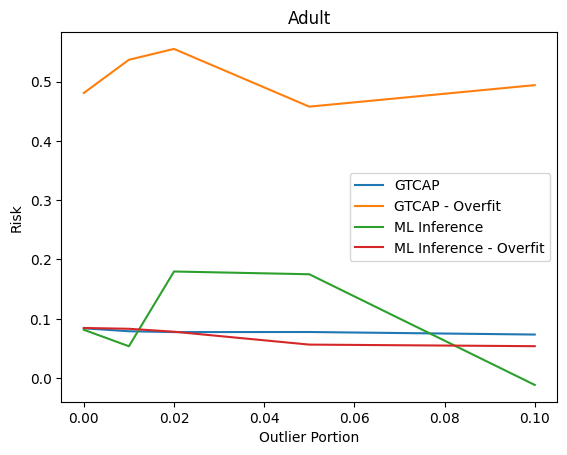}}\hfill
\subfloat
  {\includegraphics[width=.3\linewidth]{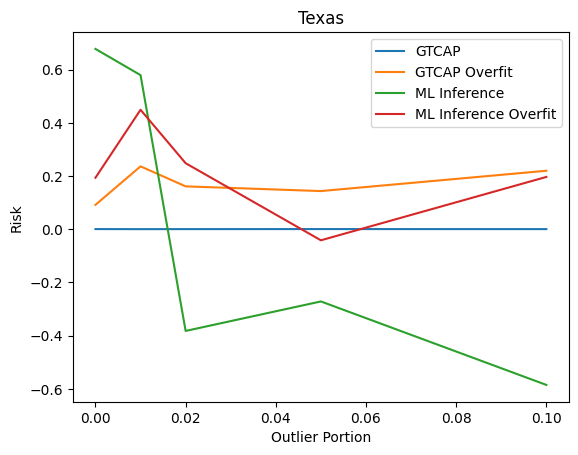}}\hfill
\subfloat
  {\includegraphics[width=.3\linewidth]{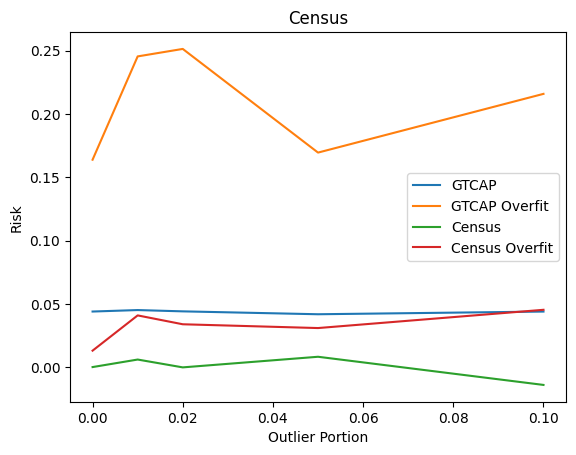}}
\caption{GTCAP and ML Inference with outlier removal in the original dataset prior to generator training, with both no overfitting ($f_o=1$) and overfitting ($f_o=1.6)$ - Synthpop~\cite{synthpop}}\label{fig:outliers-2}
\end{figure*}

\begin{figure*}[h]
\subfloat
  {\includegraphics[width=.3\linewidth]{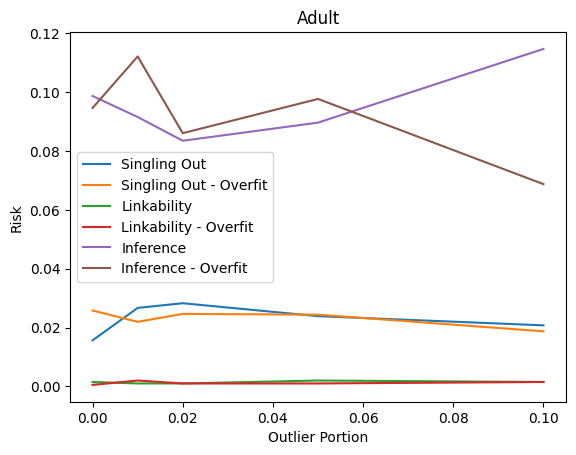}}\hfill
\subfloat
  {\includegraphics[width=.3\linewidth]{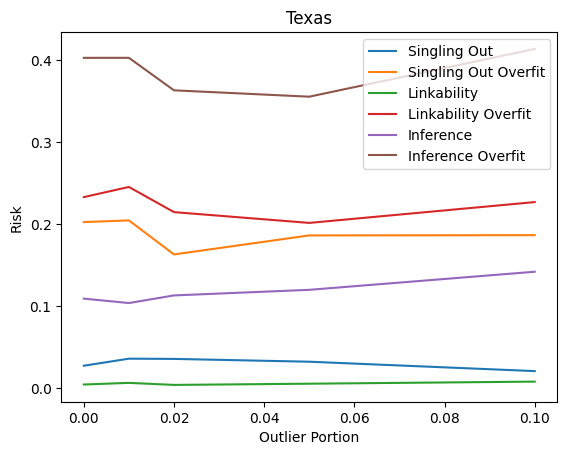}}\hfill
\subfloat
  {\includegraphics[width=.3\linewidth]{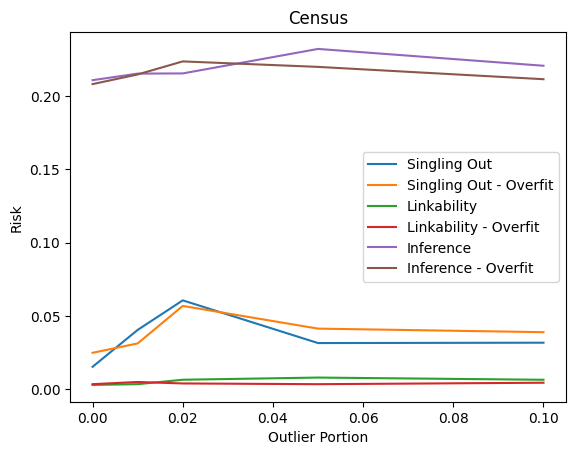}}
\caption{Anonymeter's methods with outlier removal in the original dataset prior to generator training, with both no overfitting ($f_o=1$) and overfitting ($f_o=1.6)$ - Synthpop~\cite{synthpop}}\label{fig:outliers-3}
\end{figure*}

\begin{minipage}{\textwidth}
For each metric, we applied bootstrap sampling with replacement from the original dataset ($1,000$ resamples per configuration apart from MIA and GTCAP where we resample $10$ times). Metrics were recomputed for each resample, allowing us to estimate confidence intervals. The resulting error bands correspond to 95\% confidence intervals. However, we observed that the variance across bootstrap samples is very low, typically on the order of $10^{-3}$. As a result, we decided not to plot the confidence intervals, as doing so would add visual clutter without contributing to the interpretability of the figures.
\end{minipage}




\clearpage

\section{Canary Record Baseline}\label{app:canary}

In this section, we compare the canary record baseline~\cite{TAPAS, annamalai2024linear} with the training set score for evaluating AIAs (see Section~\ref{sec:canary}) on the \emph{Adult} dataset. We repeated the baseline computations for $100$ randomly selected target records. The average AIA success rates using standard inference, machine learning-based inference attack, and GTCAP are presented below.

\begin{figure*}[h]
\subfloat
  {\includegraphics[width=.3\linewidth]{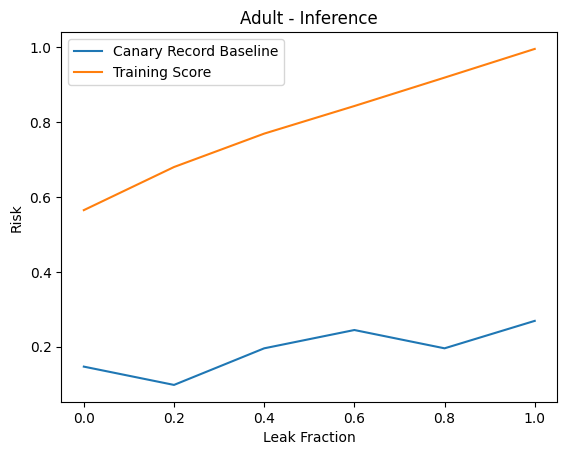}}\hfill
\subfloat
  {\includegraphics[width=.3\linewidth]{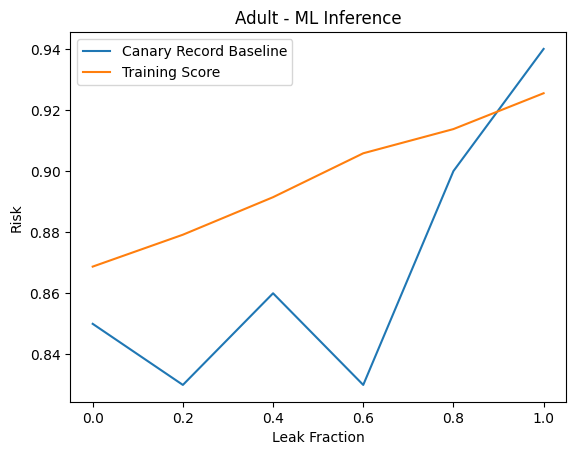}}\hfill
\subfloat
  {\includegraphics[width=.3\linewidth]{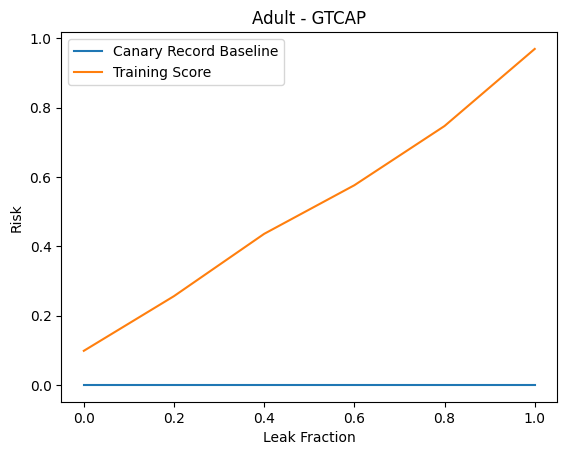}}
\caption{Comparison between Canary Record Baseline and Risk on the training set using the leaky risk model}\label{fig:MIAControl}
\end{figure*}

The Canary Record Baseline method is indeed better aligned with modern privacy auditing, particularly for synthetic data, where generalization gaps are less indicative of privacy risk. It offers a more realistic and detailed way to measure privacy risks, especially in situations where someone might actively try to extract sensitive information from synthetic data.


\section{Impact of radii on the GTCAP}\label{app:radii}

The impact of radii on GTCAP was analyzed to understand its effect on the risk measure within the training set, without using the control set as a baseline. In this experiment, a subset of attributes was selected for evaluation.

\begin{itemize}
    \item Adult dataset: The target variable was `income', while the key variables were `workclass', `education', and `marital status'.
    \item Texas dataset: The target variable was `length of stay', with key variables being `illness severity', `pat country', and `sex code'.
    \item Census dataset: The target variable was `incwage', and the key variables were `nchild', `race', and `sex'.
\end{itemize}

The results indicate that increasing the radii leads to a higher risk. This is because larger radii expand the range of values considered equivalent, thereby increasing the potential overlap and risk.

\begin{figure*}[h]
\subfloat
  {\includegraphics[width=.3\linewidth]{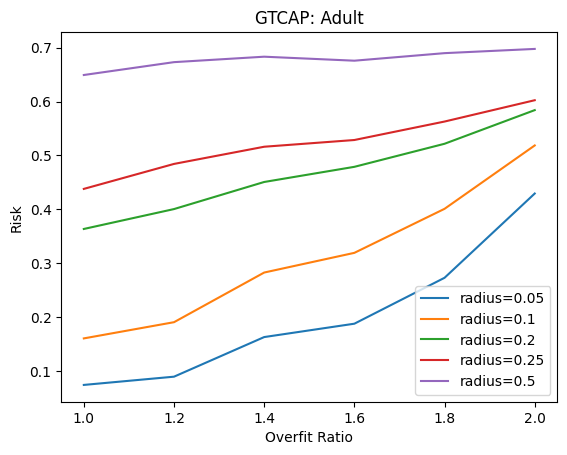}}\hfill
\subfloat
  {\includegraphics[width=.3\linewidth]{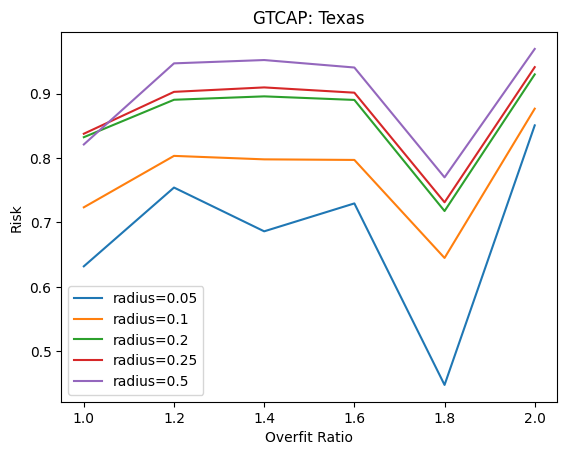}}\hfill
\subfloat
  {\includegraphics[width=.3\linewidth]{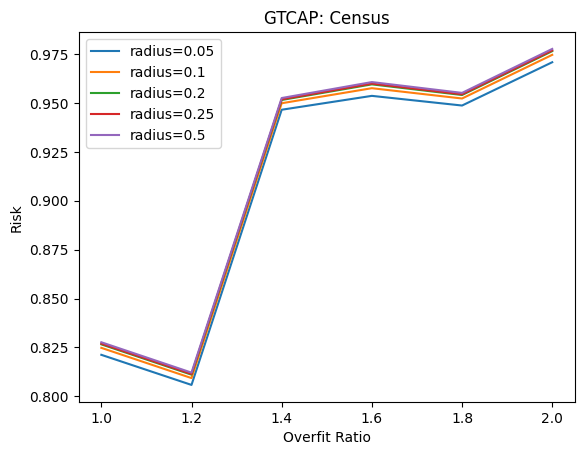}}
\caption{Results with GTCAP privacy metric for various radiuses using the overfit risk model}\label{fig:GTCAP-radius}
\end{figure*}




\section{Validation loss computation}\label{app:vallos}
Validation loss in tabular generative models—such as Realtabformer or Synthpop—is commonly used to monitor overfitting during training. These models aim to learn the joint distribution of the data, allowing them to generate realistic synthetic rows. To assess their generalization ability, we hold out a portion of the dataset as a validation set and compute the loss by measuring how well the model can reconstruct or generate these unseen rows. This involves comparing predicted values against the actual values in the held-out records. If the training loss continues to decrease while the validation loss starts increasing, it indicates that the model is memorizing specific training rows rather than learning a generalizable data distribution. This not only degrades the model's generation quality on new data but also raises the risk of privacy leakage, as it may inadvertently replicate individual records from the training set. Therefore, tracking validation loss is crucial both for assessing generalization and for ensuring the privacy-preserving nature of synthetic data generation.

\section{Utility evaluation}\label{app:util}
\begin{figure*}
\subfloat
  {\includegraphics[width=.3\linewidth]{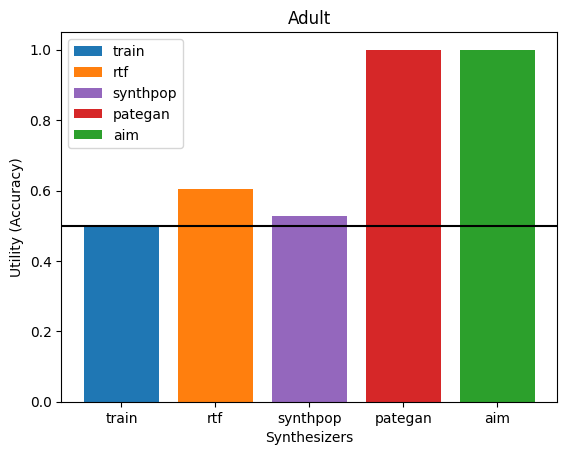}}\hfill
\subfloat
  {\includegraphics[width=.3\linewidth]{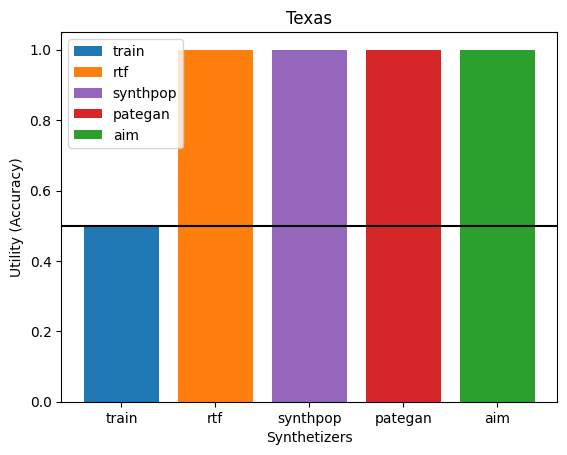}}
\subfloat
  {\includegraphics[width=.3\linewidth]{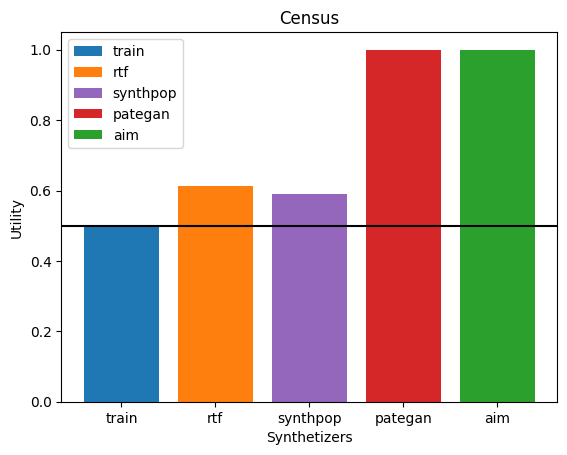}}\hfill

\caption{MLE utility scores of synthetic datasets per dataset and generator (utility of training set for reference)}\label{fig:util}
\end{figure*}

To obtain a comprehensive evaluation, we assessed the utility of the generated datasets. In particular, we measured Machine Learning Efficacy (MLE), which evaluates how well a binary classifier can distinguish between real and synthetic data samples. For this, we constructed training and test sets by combining an equal number of real and synthetic records. An XGBoost Classifier was then trained to differentiate between the two. The MLE is then the accuracy of the classifier. An MLE score close to 0.5 suggests that the synthetic data is very similar to the real data and therefore highly useful. On the other hand, a score closer to 1.0 indicates that the classifier can easily tell them apart, which implies low utility. Figure~\ref{fig:util} shows the scores per dataset and generator.

In our experiments, only REaLTabFormer and Synthpop produced synthetic data with strong utility scores. By contrast, the differentially private synthesizers scored poorly on utility. This suggests that their high privacy performance may result from generating data that simply lacks the real dataset's structure.

\section{Datasets}\label{app:data}
We use the following datasets: Adult~\cite{misc_adult_2}, the Texas Inpatient Public Use Data File (``Texas'') \cite{misc_texas}, and the 1940 Census full enumeration from IPUMS USA (``Census'')~\cite{misc_census}. We restrict the Texas and Census dataset to the following attributes:

\begin{table}[H]
\caption{Attribute selection}
\label{tab:attr}
\begin{tabular}{ll}
\toprule
 \multicolumn{1}{c}{{\bf Texas}}  & \multicolumn{1}{c}{{\bf Census}}   \\\midrule
 \small{ADMIT\_WEEKDAY, APR\_DRG, APR\_MDC, CERT\_STATUS, } & \small{AGE, AGEMARR, AGEMONTH, BPL, CHBORN, CITIZEN,}   \\
 \small{COUNTY, ETHNICITY, FIRST\_PAYMENT\_SRC, } & \small{EDUC, ELDCH, EMPSTAT, FAMSIZE, HIGRADE, HISPAN, }   \\
 \small{ILLNESS\_SEVERITY, LENGTH\_OF\_STAY, } & \small{HRSWORK1, INCWAGE, IND1950, MARST, MBPL,}  \\
 \small{PAT\_AGE, PAT\_COUNTRY, PAT\_STATE, } & \small{MOMRULE\_HIST, MTONGUE, MBPL, MOMRULE\_HIST,}\\ 
 \small{PAT\_STATUS, PAT\_ZIP, PRINC\_DIAG\_CODE, } & \small{MTONGUE, NATIVITY, NCHILD, NSIBS, OCC1950,}   \\
 \small{PROVIDER\_NAME, PUBLIC\_HEALTH\_REGION, RACE,} &  \small{POPLOC, POPRULE\_HIST, RACE, RELATE, RELATED,}  \\
 \small{RISK\_MORTALITY, SEX\_CODE, SOURCE\_OF\_ADMISSION,} & \small{SEX, SPLOC, SPRULE\_HIST, STEPMOM, STEPPOP, SUBFAM, }   \\
 \small{TOTAL\_CHARGES, TOTAL\_CHARGES\_ACCOMM, } &  \small{UOCC, WKSWORK1, YNGCH}  \\
 \small{TOTAL\_CHARGES\_ANCIL, TOTAL\_NON\_COV\_CHARGES,} &    \\
 \small{TOTAL\_NON\_COV\_CHARGES\_ACCOMM, } &    \\
 \small{TOTAL\_NON\_COV\_CHARGES\_ANCIL, TYPE\_OF\_ADMISSION} &    \\
  \bottomrule
\end{tabular}
\end{table}

The resulting datasets used in the experiments are summarized in Table~\ref{tab:data}.
        
\begin{table}[H]
\begin{center}

\caption{Datasets used in the experiments. \#Attributes: number of attributes; \#Numeric: number of numeric attributes; \#Categorical: number of categorical attributes; \#Rows: number of rows}
\label{tab:data}
\begin{tabular}{lccc}
\toprule
              & {\bf Adult}& {\bf Texas}  &  {\bf Census}   \\\midrule
\#Attributes  &   15       &      28      &  37  \\
\#Numeric     &   6        &     6        &  26  \\
\#Categorical &    9       &      21      &  11 \\
\#Rows        &     48,842 &        60,000&  75,000\\ 
\bottomrule

\end{tabular}
\end{center}
\end{table}

\newcommand{\andre}[1]{{\color{blue}[AFF: #1]}}

\end{document}